\documentclass[%
superscriptaddress,
reprint,
amsmath,amssymb,
aps,
pra,
]{revtex4-1}

\usepackage{graphicx}
\usepackage[percent]{overpic}
\usepackage{dcolumn}
\usepackage{bm}

\usepackage{hyperref}
\usepackage{color}
\definecolor{rltred}{rgb}{0.75,0,0}
\definecolor{rltgreen}{rgb}{0,0.6,0}
\definecolor{rltblue}{rgb}{0.3,0.3,1}

\usepackage{placeins}

\hypersetup{colorlinks,%
    hypertexnames=true,%
    linkcolor=rltred,%
    citecolor=rltgreen,%
    linktocpage=true}

\begin{document}
\title{Long-time expansion of a Bose-Einstein condensate: Can Anderson localization be observed?}

\author{Stefan Donsa}
\affiliation{Institute for Theoretical Physics, Vienna University of Technology,
    Wiedner Hauptstra\ss e 8-10/136, 1040 Vienna, Austria, EU}

\author{Harald Hofst\"atter}
\affiliation{Institute for Theoretical Physics, Vienna University of Technology,
    Wiedner Hauptstra\ss e 8-10/136, 1040 Vienna, Austria, EU}

\author{Othmar Koch}
\affiliation{Faculty of Mathematics, University of Vienna,
    Oskar-Morgenstern-Platz 1, 1090 Vienna, Austria, EU}

\author{Joachim Burgd\"orfer}
\affiliation{Institute for Theoretical Physics, Vienna University of Technology,
Wiedner Hauptstra\ss e 8-10/136, 1040 Vienna, Austria, EU}

\author{Iva B\v rezinov\'a}
\email{iva.brezinova@tuwien.ac.at}
\affiliation{Institute for Theoretical Physics, Vienna University of Technology,
    Wiedner Hauptstra\ss e 8-10/136, 1040 Vienna, Austria, EU}

\date{\today}

\begin{abstract}
We numerically explore the long-time expansion of a one-dimensional Bose-Einstein condensate in a disorder potential employing the Gross-Pitaevskii equation. The goal is to search for unique signatures of Anderson localization in the presence of particle-particle interactions. Using typical experimental parameters we show that the time scale for which the non-equilibrium dynamics of the interacting system begins to diverge from the non-interacting system exceeds the observation times up to now accessible in the experiment. We find evidence that the long-time evolution of the wavepacket is characterized by (sub)diffusive spreading and a growing effective localization length suggesting that interactions destroy Anderson localization.  
\end{abstract}
\maketitle

\section{Introduction}\label{sec:int}
Anderson localization in disordered systems has been predicted almost 60 years ago. Anderson could show that quantum transport is suppressed in the presence of disorder. The particle remains trapped with exponentially decaying probability density \cite{And1958}. This phenomenon turned out to be quite universal as it occurs in a wide variety of systems whenever a multitude of scattered paths can destructively interfere with each other. Localization effects have been observed in systems as diverse as electrons in solids (for a review see \cite{KraMac1993}), light in disordered media \cite{WieBarLag1997,SchBarFis2007,LahAviPoz2008}, microwaves \cite{KuhIzrKro2008}, sound waves \cite{HuStrPag2008} and in driven time-dependent systems where it is referred to as dynamical localization \cite{GreFisShm1982,GrePraFis1984}. \\
Many open questions have remained: one of the most important is the role of particle-particle interaction. In its original formulation Anderson localization has been identified within the one-particle Schr\"odinger equation. Going beyond the one-particle picture, does the many-body wavefunction of the interacting system support Anderson localization, i.e.~the exponential decay of the reduced one-body density? \\
The study of the influence of interactions on Anderson localization has a long history (e.g.~\cite{Tho1977,FleAnd80}, for a review of early results see \cite{LeeRam1985}), but has mostly been confined to time-independent condensed matter electronic systems near their ground state. The underlying many-body localization can be viewed in terms of localization in Fock space close to the Hartree-Fock state of localized single particle states \cite{BasAleAlt06}. For one-dimensional (1D) disordered systems a transition to delocalization has been found for increasing attractive interactions for fermions and increasing repulsive interactions for bosons \cite{GiaSchu1988,SchSchSch1998}. 
The ground state of the disordered Fermi-Hubbard model has been found to show different phases, a metallic phase, a Mott insulator phase, and an Anderson localized phase, depending on the interplay between the strength of disorder and on-site interactions \cite{BycHofVol05}. Similar quantum phase diagrams have been investigated for the ground state of ultracold Bose gases in 1D \cite{LugCleAsp2007}. Several theoretical (see e.g.~\cite{PalHus2010,KjaBarPol2014,KheLimShe2017}) and experimental studies (see e.g.~\cite{SchHodBor2015}) have investigated this transition from localized to delocalized many-body states in the framework of spin chains and ultracold atoms.
For finite temperature ($T> 0$) ensembles, it has been shown that at sufficiently high temperatures interactions lead to dephasing and finite conductivity \cite{AltAroKhm1982}. It was further found that electron-electron interactions yield a finite hopping above a critical temperature \cite{GorMirPol2005,BasAleAlt06}. A similar result, i.e.~a metal-insulator transition, has been found recently for weakly interacting bosons in one dimension \cite{AleAltShl2010}.\\
Recently, Bose-Einstein condensates (BECs) have offered a new platform to investigate not only Anderson localization in its pure form, i.e.~without interactions, but most importantly, in the presence of interactions \cite{SchDreKru2005,CleVarHug2005,BilJosZuo2008,RoaDerFal2008,DeiZacRoa2010,LucDeiTan2011,JenBerMul2012}. Atoms in a typical BEC interact via van der Waals interactions which can be approximated by contact interactions in the low temperature limit with the s-wave scattering length entering as the only parameter. Unless interactions are switched off by tuning Feshbach resonances in an external magnetic field, the interactions between ultracold atoms remain non-negligible \cite{RoaDerFal2008, DeiZacRoa2010}. While van der Waals or dipolar interactions are fundamentally different from the (screened) Coulomb electron-electron interaction in solids, investigation of interacting ultracold atoms in a BEC allows to probe fundamental questions about the role of interactions on Anderson localization with unprecedented experimental control with far-reaching implications also for condensed-matter physics. Numerical transport studies for 1D BECs through correlated disorder potentials showed a crossover between an Anderson localization like phase for weak interactions and a delocalized regime for larger interactions  \cite{MolTsi1994,PauLebPav2005,TiePik2008,DujEngSch2016}. The localization dynamics of time-dependent out-of-equilibrium systems has remained an open problem. Numerical simulations have been mostly performed within the framework of the discrete nonlinear Schr\"odinger equation where subdiffusive spreading of the mean width of an initially spatially localized wavepacket has been predicted and numerically observed \cite{KopKomFla2008,PikShe2008,FlaKriSko2009,LapBodKri2010,FisKriSof2012}. Closely related studies focus on the subject of diffusion or suppression of transport in out of equilibrium disordered classical chains \cite{Bas2011,DhaLeb2008,OgaPalHus2009}.\\
In the present paper, we aim for a numerical simulation following the experiment \cite{BilJosZuo2008} in which first indicators of Anderson localization in an expanding BEC in one dimension were reported. The goal is to explore the long-time evolution of such a non-equilibrium system and to identify possible signatures of Anderson localization. This requires the solution of the continuous rather than the discrete non-linear Schr\"odinger equation whose long-time propagation poses a considerable numerical challenge. In the experiment \cite{BilJosZuo2008}, the BEC is initially confined in a cigar-shaped harmonic trap. The longitudinal harmonic potential is switched off after which expansion in a speckle-like disorder potential along the longitudinal axis sets in. Observation of an approximately exponentially localized particle density of almost macroscopic extent and over a period of more than one second after switching off the longitudinal trapping potential provided first indication of Anderson localization of an expanded BEC. \\
We solve the Gross-Pitaevskii equation (GPE) on a large spatial grid and for long propagation times to explore the effect of particle-particle interactions on the expansion. In particular, we delineate the intrinsic difficulties in identifying unique signatures of Anderson localization within such a non-equilibrium scenario resulting from the simultaneous presence of multiple time and length scales in this problem. \\
We relate our findings to the destruction of Anderson localization by subdiffusive spreading where the mean width increases as $\Delta x \propto t^{\alpha}$ as a function of time $t$ with $\alpha < 1/2$ as observed in discrete systems \cite{FlaKriSko2009,LapBodKri2010}. We show that both a clear distinction between the expansion of an interacting and a non-interacting BEC as well as the approach to an asymptotically localized state, if occurring at all, manifests itself only on time scales exceeding present experiments.\\
The paper is structured as follows: In Sec.~\ref{sec:obs_num} we introduce the system under consideration, the observables, and the numerical method used to propagate the system. In Sec.~\ref{sec:sce} we introduce the alternative expansion scenarios in order to disentangle the roles of particle-particle interactions and disorder. Results for observables signifying the degree of localization will be presented in Sec.~\ref{sec:macobs} followed by conclusions in Sec.~\ref{sec:sum}. Further technical details are given in the appendices \ref{App:numerics} to \ref{App:theory}.
\section{System, and numerical implementation}\label{sec:obs_num}
We simulate a typical expansion scenario for a BEC following the experimental scenario of Ref~\cite{BilJosZuo2008}. Accordingly, we consider a BEC  of $^{87}$Rb atoms trapped in a cigar-shaped trap with transverse frequency $\omega_\perp=70\times 2\pi$~Hz and longitudinal frequency $\omega_0=5.4\times 2\pi$~Hz. Inter-atomic interactions are governed by the s-wave scattering length which is $a_s=5.8$nm \cite{BurBohEsr1998}. The characteristic scales for length, time, and energy of the initial longitudinal harmonic trap potential serve as units throughout the paper, i.e.~$x_0=\sqrt{\hbar/m\omega_0}\approx4.6\mu$m, $t_0=2\pi/\omega_0\approx29$ms, and $e_0=\hbar\omega_0$ ($\hbar=m=\omega_0=1$). The BEC wavefunction along the longitudinal axis $x$ is then given the solution of the 1D stationary GPE
\begin{align}\label{eq:gs_GPE}
	-\frac{\hbar^2}{2m}\frac{\partial^2}{\partial x^2}\psi(x) 
	+ \frac{m\omega_0^2x^2}{2}\psi(x)& \nonumber \\
	+ 2\hbar\omega_\perp a_sN|\psi(x)|^2\psi(x) &= \mu_0\psi(x),
\end{align}
where $N$ is the particle number and $\mu_0$ is the chemical potential. For simplicity we abbreviate the strength of the inter-particle interaction by the nonlinearity-parameter $g=2\hbar\omega_\perp a_sN$. For the numerical value of the nonlinearity we use $g_0=400e_0x_0$ which corresponds to a particle number of $N\approx1.2\times10^4$. Calculating the ground state with Eq.~\ref{eq:gs_GPE} one obtains for the chemical potential $\mu_0 \approx35.6e_0$ which is slightly larger than the energy of the first transverse excited state in the trap $e_\perp^1=2\hbar\omega_\perp+\hbar\omega_0/2=26.4e_0$ indicating that the radial degree of freedom might not be completely negligible in the experiment, at least initially. As discussed below, the BEC released from the longitudinal trap, while keeping the radial trap intact, rapidly expands such that the chemical potential falls below $e_\perp^1$ within $t_0$. We therefore assume (as has been suggested in \cite{BilJosZuo2008}) that the ensuing dynamics can be regarded as effectively one-dimensional. The solution to Eq.~\ref{eq:gs_GPE} provides the initial condition at $t=0$ for the subsequent expansion dynamics. For $t>0$ we describe the dynamics within the 1D time-dependent GPE as
\begin{align}\label{eq:td_GPE}
	-\frac{\hbar^2}{2m}\frac{\partial^2}{\partial x^2}\psi(x,t) 
	+ V(x,t)\psi(x,t)& \nonumber \\
	+ g(t)|\psi(x,t)|^2\psi(x,t) &= i\hbar \frac{\partial}{\partial t}\psi(x,t).
\end{align}
Under the condition $r_s\gg a_s$, where $r_s$ is the mean distance between the particles, the many-body ground state in the trap corresponds to a BEC well described by the GPE \cite{PetShlWal2000}.
In Eq.~\ref{eq:td_GPE}, the harmonic trapping potential in Eq.~\ref{eq:gs_GPE} is switched off at $t=0$ and the disorder (D) potential 
\begin{align}\label{eq:V(x,t)}
	V(x,t)=V_D(x)\Theta(t-t_D)
\end{align}
is switched on at the time $t_D > 0$. We also allow for a time-dependent interaction strength
\begin{align}\label{eq:g(t)}
	g(t)=g_0\left[1-\Theta(t-t_I) \right],
\end{align}
whereby the particle-particle interactions, represented in the GPE by the non-linearity, are switched off at $t_I>0$. Choosing suitable $t_D$ and $t_I$ allows us to explore the competition between disorder and particle-particle interactions during the expansion process.\\
For the disorder potential we consider both Gaussian correlated disorder potentials and speckle potentials \cite{Hun1989,HorCouGry1998}. 
\begin{figure}[t]
	\includegraphics[width = 0.9\columnwidth]{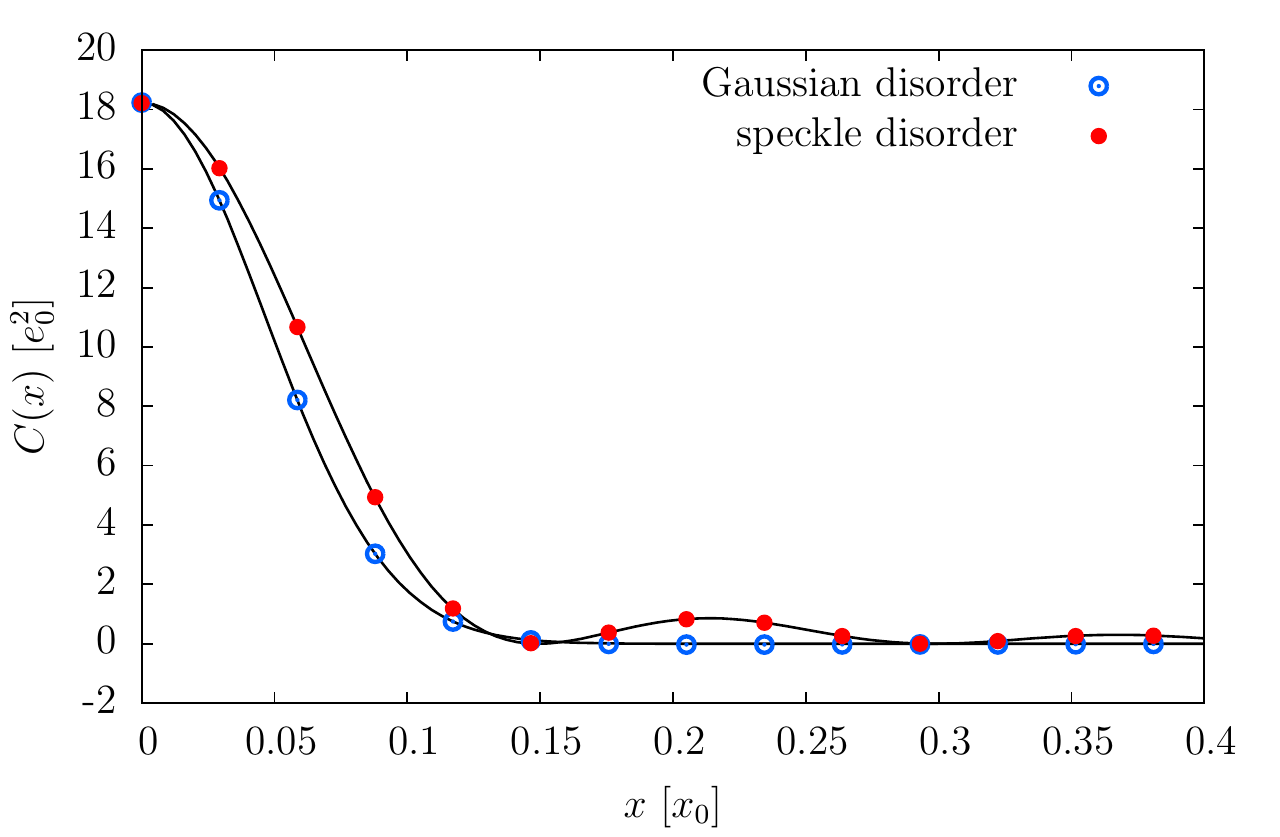}
	\caption{(Color online) Real-space correlation function for Gaussian correlated (Eq.~\ref{eq:corr_G}) and speckle disorder (Eq.~\ref{eq:corr_speck}).}
	\label{fig:corr_pot}
\end{figure}
The Gaussian correlated disorder potential is generated by equidistantly placing Gaussian functions of width $\sigma_D$ and random amplitude $A_i$
\begin{align}
	U(x) = \sum_{i=1}^{N_d}A_ie^{\frac{-(x-x_i)^2}{2\sigma_D^2}}.
\end{align}
The $A_i$'s are uniformly distributed in the interval $[-1,1]$. The distance between adjacent Gaussian functions is $\delta x=x_{i+1}-x_i=5\times 10^{-4}x_0$. After subtracting the mean, $\tilde{U}(x)=U(x)-\left< U(x)\right>$, and normalizing by the standard deviation, $\left< \tilde{U}(x)^2 \right>^{1/2}$, the disorder potential becomes
\begin{align}
	V(x)=\frac{V_D}{\left< \tilde{U}(x)^2 \right>^{1/2}}\tilde{U}(x).
\end{align}
To be in the regime of Anderson localization \cite{SanCleLug2007} we choose for $\sigma_D=0.39\xi$ where $\xi=\sqrt{\hbar/4m\mu_0}$ is the healing length. For $V_D$ we take $V_D=0.3[E_{\rm int}(t=0)+E_{\rm kin}(t=0)]$. The disorder is characterized by a Gaussian correlation function in real space (Fig.~\ref{fig:corr_pot})
\begin{align}\label{eq:corr_G}
	C_G(x)=V_D^2e^{\frac{-x^2}{\left( 2\sigma_D \right)^2}},
\end{align}
and a Gaussian correlation function
\begin{align}\label{eq:k_corr_G}
	\tilde C_G(k)&=\frac{1}{\sqrt{2\pi}}\int_{-\infty}^{\infty} {\rm d}x\; C_G(x)e^{-ikx} \nonumber \\
	&=\sqrt{2}\sigma_DV_D^2e^{-k^2\sigma_D^2}.
\end{align}
also in Fourier space. For the speckle potential we follow the procedure described in Ref.~\cite{Hun1989,HorCouGry1998}. The speckle potential is characterized by the real-space correlation function
\begin{align}\label{eq:corr_speck}
	C_S(x)=V_D^2\frac{\sin(x/\sigma_D)^2}{(x/\sigma_D)^2},
\end{align}
whose Fourier transform 
\begin{align}\label{eq:k_corr_speck}
	\tilde C_S(k)=\sqrt{\pi/2}\sigma_D V_D^2\left(1-\frac{\left|k\right| \sigma_D}{2}\right)
	\Theta\left(1-\frac{\left| k \right| \sigma_D}{2}\right),
\end{align}
has a sharp high-momentum cut-off. We have chosen the parameters for the speckle potential such that the correlation functions in Fourier space have the same value at the maximal momentum $k_{\mathrm{max}}=1/\xi$ \cite{SanCleLug2007}  within the Thomas-Fermi approximation 
\begin{align}\label{eq:par_speckle}
	\tilde C_S(k=k_{\mathrm{max}})=\tilde C_G(k=k_{\mathrm{max}}).
\end{align}
The numerical values used for the speckle potential are $V_D=0.3[E_{\mathrm{int}}(t=0)+E_{\mathrm{kin}}(t=0)]$ and $\sigma_D=0.57\xi$.  As will be discussed below, the choice of a (relatively) small value of $\sigma_D$ facilitates the observation of differences between the interacting and non-interacting scenarios already at (relatively) short propagation times. The choice of parameters [Eq.~\eqref{eq:par_speckle}] leads to somewhat different real-space correlation functions for the two disorder potentials (see Fig.~\ref{fig:corr_pot}). We have, however, verified that choosing different parameters for the speckle disorder such that the real-space correlation functions would more closely resemble each other does not significantly alter our results. \\
The observables we will be focusing on are the disorder-averaged real-space particle density
\begin{align}\label{eq:density}
	n(x,t)=\langle|\psi(x,t)|^2\rangle_{\rm disorder},
\end{align}
and momentum space density
\begin{align}\label{eq:mom_density}
	n(k,t)=\langle|\tilde\psi(k,t)|^2\rangle_{\rm disorder}
\end{align}
with
\begin{align}\label{eq:FT_wfn}
	\tilde\psi(k,t)=\frac{1}{\sqrt{2\pi}}\int_{-\infty}^{\infty} {\rm d}x\,e^{-ikx}\psi(x,t).
\end{align}
The disorder-averaged particle densities in either coordinate or momentum space displayed in the following are smoothed with Gaussians of width $6 x_0$ and $0.1/x_0$, respectively. For the analysis of (sub)diffusive spreading we investigate the mean width in real space 
\begin{align}\label{eq:delta_x}
	\Delta x(t) = \sqrt{\langle x^2 \rangle - \langle x \rangle^2}, 
\end{align}
with
\begin{align}\label{eq:xn}
	\langle x^n\rangle=\left\langle \int {\rm d}x\; |\psi(x,t)|^2 x^n\right\rangle_{\rm disorder},	
\end{align}
and the corresponding width in Fourier space
\begin{align}\label{eq:delta_k}
	\Delta k(t) = \sqrt{\langle k^2 \rangle - \langle k \rangle^2}.
\end{align}
As will be discussed below, due to the multi-scale nature of the expansion scenario, the mean width is insufficient to completely characterize the localization process. We therefore define, in addition, the inverse slope of the decaying density distribution $L_{\rm loc}$ as a measure for the degree of localization (see below). \\
For the numerical propagation we discretize space in an equidistant grid and use a spectral (Fourier) basis to represent spatial derivatives. A sufficiently dense grid was chosen such that the results have converged with respect to the grid spacing. This implies that we are in true continuum limit and are not treating a discrete non-linear Schr\"odinger equation as has been investigated in the past \cite{KopKomFla2008,PikShe2008,FlaKriSko2009,LapBodKri2010}. For all results presented here we use a box size $x \in \left[ -2.4 \times 10^4, 2.4 \times 10^4 \right]x_0$ with 1638400 grid points and periodic boundary conditions. For the time propagation we use a split-step method with adaptive step size.  
Details of the numerical method allowing to reach long propagation times are given in the Appendices \ref{App:numerics} and \ref{App:convergence}. \\
The validity of the GPE for describing the non-equilibrium dynamics of this many-body body system is a priori not obvious. However, previous investigations focusing on disorder averaged observables such as $\Delta x$ have shown good agreement with advanced many-body treatment such as the multi-configurational time-dependent Hartree method for bosons (MCTDHB) \cite{BreColLud2011,BreLodStr2012,AloStrCed2008} at least for short times.
\section{Scenarios for the expansion}\label{sec:sce}
%
\begin{figure}[t]
	\includegraphics[width = 0.9\columnwidth]{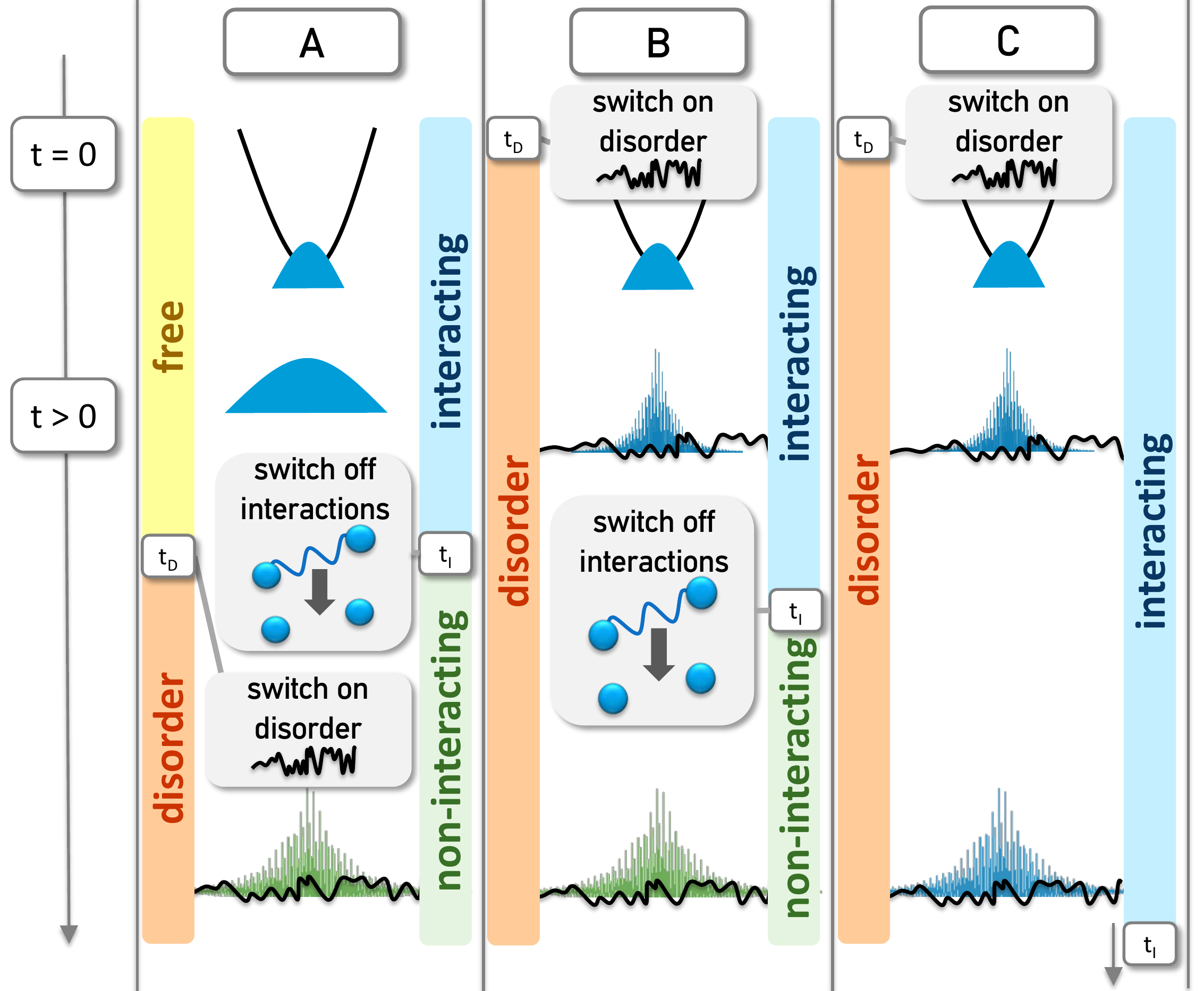}
	\caption{(Color online) The three scenarios discussed in this paper: scenario A (left column), scenario B (central column), and scenario C (right column). For details see text.}
	\label{fig:scen}
\end{figure}
We consider now three different scenarios for the expansion of a BEC in a disorder potential (Fig.~\ref{fig:scen}). They allow to disentangle the influence of particle-particle interactions and disorder on the non-equilibrium dynamics and localization of the wavepacket during the expansion. In scenario A, we let the BEC freely expand up to $t=t_D$. The switch-on time of the disorder $t_D$ is chosen such that the initial particle-particle interaction energy 
\begin{align}
	E_{\mathrm{int}}=\frac{1}{2}g \int \mathrm{d} x \left|\psi(x,0) \right|^4
\end{align} 
is almost completely converted into kinetic energy. After the time $t_D$ we also switch off interactions, i.e.~we set $t_D=t_I$ in Eq.~(\ref{eq:V(x,t)}) and Eq.~(\ref{eq:g(t)}). As numerical value we choose $t_D=20 t_0$. At this point the instantaneous interaction energy $E_{\mathrm{int}}(t)$ is reduced to $E_{\mathrm{int}}(t)=0.04E_{\mathrm{total}}$. In scenario B, the disorder potential is present during the entire expansion time (i.e.~$t_D=0$) but the particle-particle interactions (or nonlinearity) are switched off at $t_I=20 t_0$. In this case, $E_{\mathrm{int}}(t)=0.25E_{\mathrm{total}}$, i.e.~a significantly larger fraction of the total energy is neglected at the point of switch-off. Finally, in scenario C the disorder is also present from the start of the expansion ($t_D=0$) as in scenario B but the particle-particle interactions are never switched off during the entire simulation (i.e. $t_I \rightarrow \infty$). It is the latter scenario that most closely represents the experiment \cite{BilJosZuo2008}.
\subsection{Scenario A}\label{sec:scen_A}
\begin{figure}[t]
	\includegraphics[width = 0.9\columnwidth]{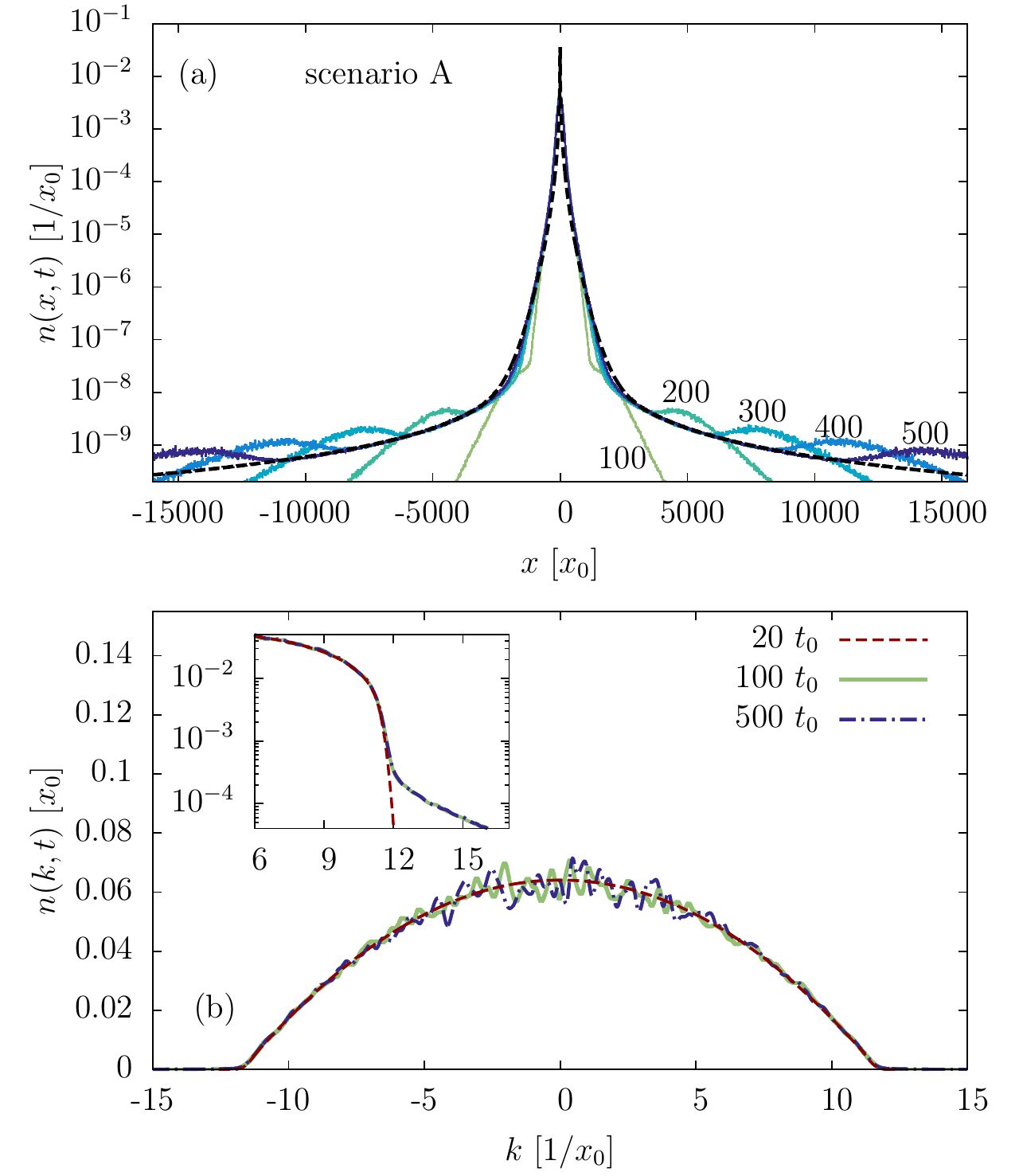}
	\caption{(Color online) (a) Numerically obtained densities for scenario A (see Fig.~\ref{fig:scen}) with the Gaussian correlated disorder for several time steps indicated by the numbers in units of $t_0$ next to the individual curves. The densities have been calculated within a disorder average over 50 ensembles. The analytical formula, Eq.~\ref{eq:den_analyt}, is plotted as a reference (black dashed line). (b) The momentum distribution for two later times as compared to the initial momentum distribution for $t_I=20t_0$. The same ensemble as in (a) has been used. Inset: The momentum distribution in log-scale.}
	\label{fig:wfn_A}
\end{figure}

This scenario has been previously used to describe analytically \cite{SanCleLug2007} the experimental observation in Ref.~\cite{BilJosZuo2008}. The underlying assumption is that during the initial expansion stage (up to $t=t_D$) where most of the interaction energy is transferred into kinetic energy  the disorder can be neglected while for later $t>t_D=t_I$ the remaining interaction energy can be neglected. In the present case switching off the interaction, i.e.~setting $g=0$, at $t_D=t_I$ amounts to neglecting approximately $4\%$ of the total energy $E_{\rm total}$. Snapshots of the simulated evolved real-space and momentum density are shown in Fig.~\ref{fig:wfn_A}. \\ 
The analytical description \cite{SanCleLug2007,SanCleLug2008,PirLugBou2011,ValWel2016} of the emerging asymptotic real-space density for this scenario
\begin{align}
	n(x)=\lim_{t\rightarrow \infty}n(x,t),
\end{align}
uses the fact that each momentum space component, i.e.~plane wave of the wavepacket amplitude $\tilde{\psi}(k,t_I)$ in the by now non-interacting system localizes with a $k$ (or energy) dependent localization length 
\begin{align}\label{eq:gamma}
	\lambda [E(k)]=\sqrt{\frac{2}{\pi}}\frac{k^2}{\tilde{C}(2k)},
\end{align}
with $E=k^2/2$. \\
Eq.~\ref{eq:gamma} follows from the first Born approximation which applies to high energies and weak disorder, i.e. $E\gg\sqrt{ \left< V_D^2  \right>}$ and $V_D \sigma_D \ll \frac{\hbar^2 k}{m} \left(k \sigma_D\right)^{1/2} $. Consequently, the localization length diverges in the limit of large energies in this approximation. For a Gaussian disorder potential $\lambda$ grows exponentially with $E$, for a speckle potential it diverges for $E > \frac{2}{\sigma_D^2}$. As the expanding BEC acquires a high-energy tail during the conversion of $E_{\mathrm{int}}$ into kinetic energy, a fraction of the BEC executes a rapid quasi-ballistic expansion thereby counteracting localization. The quasi-ballistic expansion is clearly seen in the rapidly moving tails of the real space density [Fig.~\ref{fig:wfn_A} (a)]. \\
The asymptotic density distribution can be derived from
\begin{align}\label{eq:den_analyt}
	n(x)=\int {\rm d}k \int {\rm d}E\, |\tilde \psi(k,t_I)|^2 A(k,E)\,n[\lambda(E),x],
\end{align}
where $n[\lambda(E),x]$ is the density of the Anderson localized state with localization length $\lambda (E)$ \cite{PirLugBou2011,ValWel2016}.  $A(k,E)$ denotes the spectral function (see Appendix \ref{App:theory}) and $\left|\tilde{\psi}(k,t_I)\right|^2$ is the initial momentum distribution.  In this case the ``initial" state of the localization scenario is the momentum distribution at the time the particle-particle interaction is switched off. The wavepacket at $t=t_I$ can be calculated analytically as \cite{KagSurShl1996,SanCleLug2008} 
\begin{align}\label{eq:momentum_den_analyt}
	|\tilde{\psi}(k,t_I)|^2 = \frac{3\xi}{4}\left(1-k^2\xi^2\right)\Theta(1-|k|\xi),
\end{align}
shown in Fig.~\ref{fig:wfn_A} (b) where we have chosen $t_I$ such that the stationary momentum distribution [Eq.~\eqref{eq:momentum_den_analyt}] is reached to a high degree of accuracy. Eq.~[\ref{eq:den_analyt}] predicts very flat tails reaching out to infinity [Fig.~\ref{fig:wfn_A} (a)]. 
\begin{figure}[t]
	\includegraphics[width = 0.9\columnwidth]{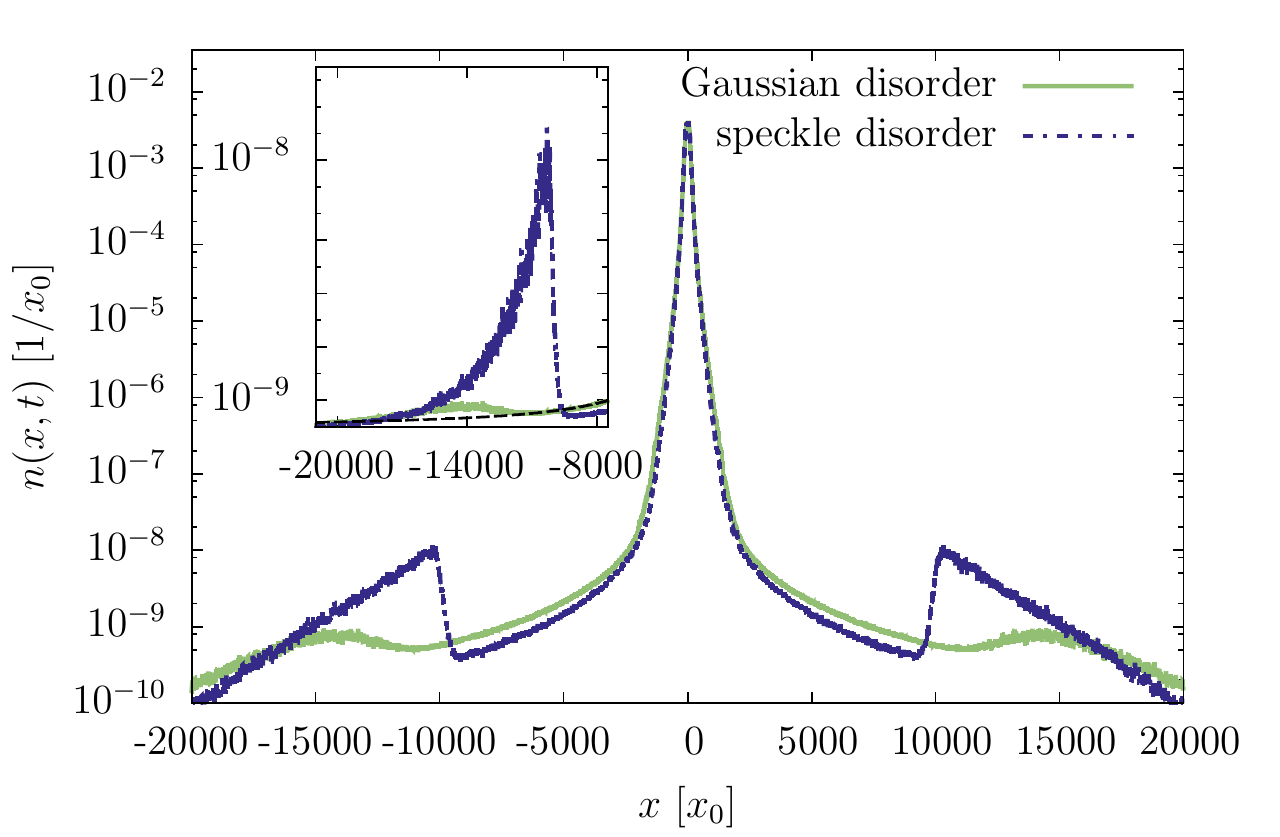}
	\caption{(Color online) Comparison between the densities at $t=500t_0$ for the Gaussian correlated disorder and for the speckle disorder. Inset: Linear plot of the densities at the position of the local maxima. Eq.~\ref{eq:den_analyt} plotted as reference (black dashed).}
	\label{fig:wfn_A_gauss_speckle}
\end{figure}
Comparing with the numerically calculated densities as a function of time we observe very good agreement for $t> t_I$. No fitting parameter is involved here. The numerically obtained non-stationary densities feature tails with local maxima above the analytical prediction. The local peaks move ballistically while at the same time successively spreading out. These tails contain high momentum components that only weakly scatter at the disorder potential. In fact, for the speckle potential, the scattering probability vanishes. Consequently, the local peaks are even more pronounced for the speckle potential than for the Gaussian correlated disorder (Fig.~\ref{fig:wfn_A_gauss_speckle}). Obviously, taking the width $\Delta x$ of the entire wavepacket as measure for the localization fails to capture the multi-scale nature of the expansion process that features simultaneously quasi-ballistic expansion of the outer tails and a fairly localized central region which may, possibly, display (sub)diffusive spreading.\\
It is worth noting that the ``initial" moment distribution $\left|\tilde{\psi}(k,t_I) \right|^2$ at the moment of switching-off the inter-particle interaction [Eq.~\eqref{eq:momentum_den_analyt}] is only weakly perturbed by the subsequent propagation in the disorder [Fig.~\ref{fig:wfn_A} (b)].  This, at first sight, surprising result can be explained as follows. For each disorder realization the wavepacket can be expanded in Anderson modes, the eigenstates of the linear Schr\"odinger equation with the  disorder potential, as
\begin{align}\label{eq:Anderson_modes}
	\tilde{\psi}(k,t) = \sum_n a_n(t_I) e^{-iE_n (t-t_I)}\phi_n(k),
\end{align}
where $\phi_n(k)$ are the Anderson modes in momentum space and $a_n$ are the expansion coefficients at time $t_I$. Note that Eq.~\eqref{eq:Anderson_modes} is valid only in the non-interacting regime of scenario A and B [i.e. $g(t)=0$ for $t>t_I$]. The momentum density is then given by
\begin{align}\label{eq:momentum_den_Anderson_modes}
	|\tilde{\psi}(k,t)|^2 &= \sum_n |a_n|^2 |\phi_n(k)|^2 \nonumber \\
	& + \sum_{n\neq m}a_m^*a_n e^{i(E_m-E_n)(t-t_I)}\phi_m^*(k)\phi_n(k).
\end{align}
For large differences in energy $|E_m-E_n|$, the Anderson modes will have a small overlap with each other in momentum space such that the corresponding off-diagonal contributions in the sum can be neglected. For small (but non-vanishing) differences $|E_m-E_n|$ and large $E_m$, $E_n$ the overlap still will be negligible since the width of the Anderson modes in momentum space
\begin{align}
	\phi_n(k) \propto \left[1+\left[ (k-k_n) \lambda (E_n) \right]^2  \right]^{-1}
\end{align}
rapidly decreases. The approximate Lorentzian shape of $\phi_n(k)$ follows from the fact that for large energies the Anderson modes in real space closely resemble plane waves with an exponentially localized envelope $~\exp\left[-\left|x\right| / \lambda(E)\right]$. Perturbative corrections to $\phi_n(k)$ are neglected. We consistently observe the smallest fluctuations [i.e. deviations from $\left|\tilde\psi(k,t_I)\right|^2$] for large momenta. The disorder average further reduces the off-diagonal terms such that we observe very good agreement between $n(k,t_I)$ and $n(k,t)$ for $t=100t_0$ and $t=500t_0$, see Fig.~\ref{fig:wfn_A}~(b). The tails [see the log-plot in the inset of Fig.~\ref{fig:wfn_A} (b)] represent the momentum distribution of the Anderson modes, i.e. the diagonal contributions to Eq.~\eqref{eq:momentum_den_Anderson_modes}.
\subsection{Scenario B}\label{sec:scen_B}
%
\begin{figure}[t]
	\includegraphics[width = 0.9\columnwidth]{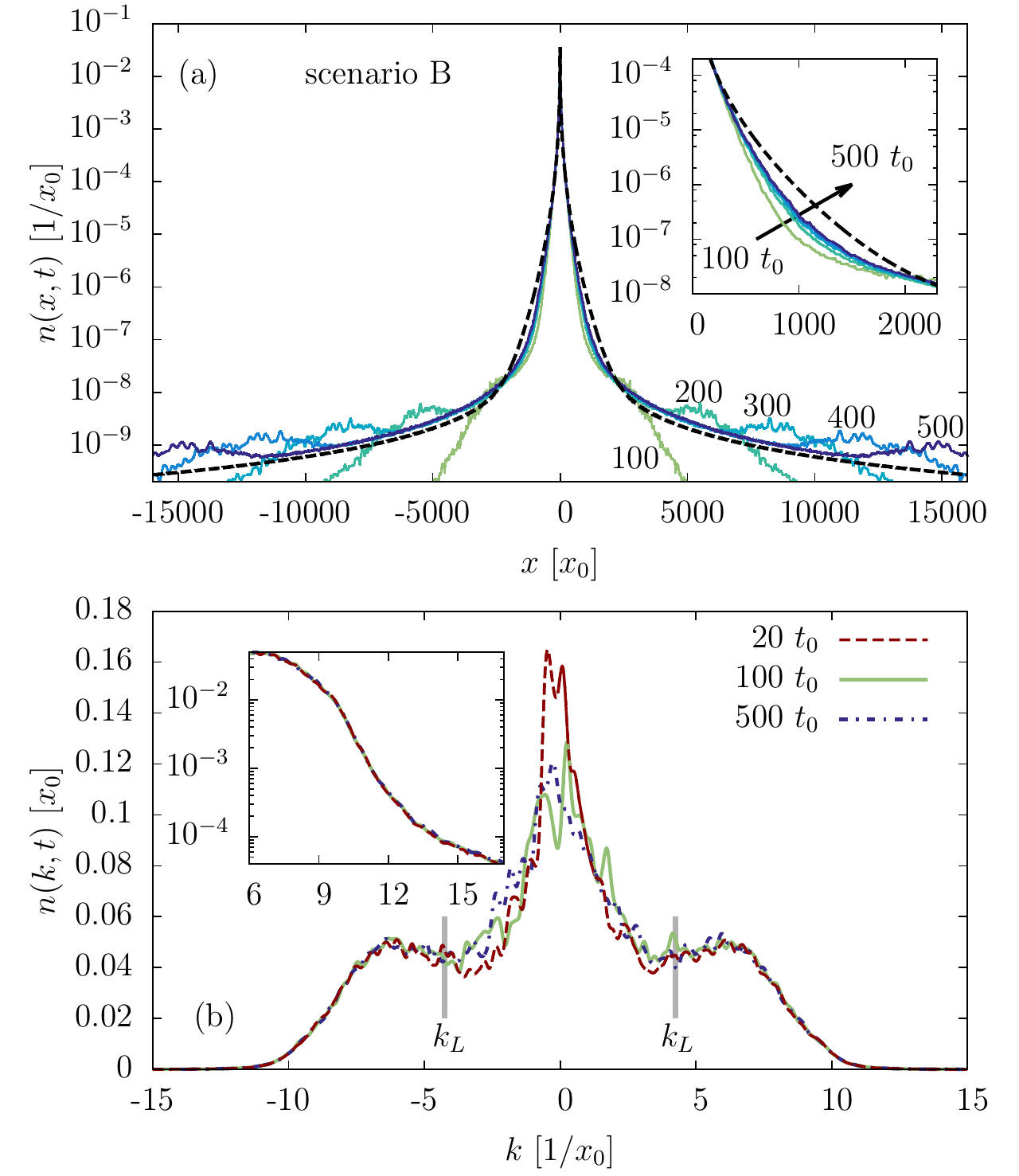}
	\caption{(Color online) (a) Numerically obtained densities for scenario B (see Fig.~\ref{fig:scen}) for the Gaussian correlated disorder at several time steps indicated by the numbers in units of $t_0$ next to the individual curves. The densities have been calculated within a disorder average over 50 ensembles. The analytical formula, Eq.~\ref{eq:den_analyt} black dashed line, plotted as reference. (b) The momentum distribution for two later times as compared to the initial momentum distribution at $t_I=20t_0$. The same ensemble as in (a) has been used. The vertical lines mark the position of $k_L=\pm \sqrt{\mu(t_s)}$. Inset: The momentum distribution in log-scale.}
	\label{fig:wfn_B}
\end{figure}
This scenario distinguishes itself from scenario A by the immediate presence of the disorder potential from $t=0$ on ($t_D=0$). As in scenario A, the particle-particle interactions are switched off at $t_I$, $\left[g(t>t_I)=0\right]$. This scenario allows to investigate the influence of disorder on the initial expansion process, in particular, on the momentum distribution that provides the initial condition for the subsequent non-interacting Anderson scenario for $t> t_I$. The presence of disorder during the time interval $0<t<t_I$ gives rise to new features not present in scenario A: In presence of the disorder potential the conversion of interaction energy into kinetic energy cannot proceed unhindered. Instead, the initial superfluid flow experiences friction in form of excitation of phonons as soon as momenta above the Landau velocity (or momentum) $k_L$ become available \cite{PetSmi2008}. Those momenta then scatter inelastically at the disorder potential. Correspondingly, we observe for $t<t_I$ the emergence of minima in the momentum distribution around $\pm k_L$ \cite{BreLodStr2012} where $k_L=\sqrt{\mu(t_I)}$, and an enhanced density for smaller momenta [Fig.~\ref{fig:wfn_B} (b)]. Once the pair interaction is switched off for $t>t_I$, the disorder averaged momentum density remains essentially unchanged as was the case in scenario A. It should be noted that in scenario B the sudden switch-off of the interaction amounts to neglecting quite a large fraction ($\sim25\%$) of the initial total energy $E_{\mathrm{tot}}$ (compared to 4 \% in scenario A) reflecting the suppression of energy conversion by disorder. Unlike scenario A, scenario B cannot be treated fully analytically because the momentum distribution at $t=t_I$ is not known analytically. However, we observe only little deviations in the numerically obtained time-dependent densities as compared to scenario A confirming that the modified initial momentum distribution has indeed a minor effect on the shape of the densities at later times as assumed in \cite{SanCleLug2007,SanCleLug2008, BilJosZuo2008} [Fig.~\ref{fig:wfn_B} (a)]. Likewise, the present results suggest that inserting the numerically found momentum distribution at $t=t_I$, $\left|\tilde{\psi} (k,t_I) \right|^2$, in Eq.~\ref{eq:den_analyt} allows for a reasonably accurate description of the asymptotic real-space density $n(x)$.
\subsection{Scenario C}\label{sec:scen_C}
%
\begin{figure}[t]
	\includegraphics[width = 0.9\columnwidth]{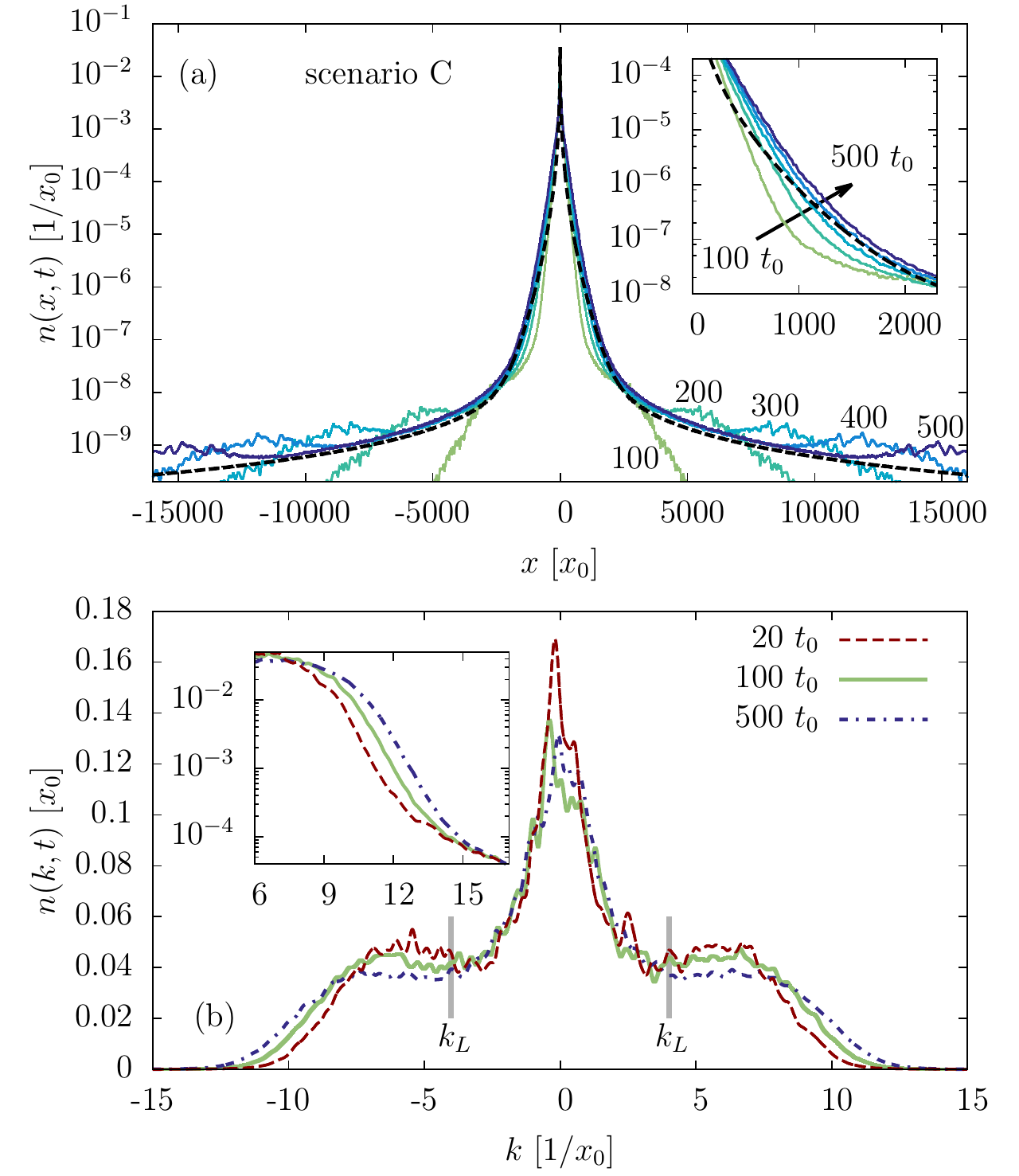}
	\caption{(Color online) (a) Numerically obtained densities for scenario C (see Fig.~\ref{fig:scen}) for the Gaussian correlated disorder at several time steps indicated by the numbers in units of $t_0$ next to the individual curves. The densities have been calculated within an ensemble average over 50 ensembles. The analytical formula, Eq.~\ref{eq:den_analyt} black dashed line, plotted as reference. Inset: zoom-in of the expanding density for several time steps, starting with $t=100t_0$ up to $t=500t_0$ (order indicated by the arrow). (b) The momentum distribution for two later times as compared to the initial momentum distribution for $t_I=20t_0$. The same ensemble as in (a) has been used. The vertical lines mark the position of $k_L=\pm \sqrt{\mu(t)}$ where the width of the lines corresponds to the standard deviation due to the time dependence of $\mu(t)$ and averaging over ensembles. Inset: The momentum distribution in log-scale.}
	\label{fig:wfn_C}
\end{figure}
Within the mean-field approximation, scenario C is the one which should most closely describe the experimental situation: Disorder is present from the start and particle-particle interactions are present throughout the propagation. Moreover, comparing scenario C with B which differ only by the presence (or absence) of interactions beyond $t=t_I$ allows to identify the effect of interactions on Anderson localization for long times. \\
At first glance, the time-dependent density $n(x,t)$ for scenario C (Fig.~\ref{fig:wfn_C}) closely resembles that of the two previous scenarios A and B. The density features simultaneously a strongly peaked region and rapidly ballistically moving tails. There is, however, a unique feature only present in this case (see also \cite{MinLiRos2012}): The width of the central region ($x\in [-2000x_0,2000x_0]$) is continuously expanding [Fig.~\ref{fig:wfn_C} (a)] while it becomes stationary in scenario $B$ at approximately $t=200t_0$ [Fig.~\ref{fig:wfn_B} (a)]. The effect can also be observed in the momentum distribution which continuously broadens as a function of time [Fig.~\ref{fig:wfn_C} (b)]. It is important to note that detecting this feature requires disorder averages with sufficiently large ensemble sizes to prevent blurring the signal by large fluctuations present. Observing them in the experiment may pose a considerable challenge as both a sufficient spatial resolution of the density as well as a large enough ensemble would be needed.\\
In presence of particle-particle interactions Anderson modes are no longer eigenstates of the full Hamiltonian. Consequently, the expansion of the wavepacket in Anderson modes features now time-dependent expansion coefficients 
\begin{align}
	\psi(x,t) = \sum_n a_n(t)e^{-iE_nt}\phi_n(x),
\end{align}
whose time evolution is governed by
\begin{align}\label{eq:td_Anderson_modes}
	i\partial_t a_n(t) &= g\sum_{m_1,m_2,m_3}V_{m_1,m_2}^{n,m_3}a_{m_3}^*(t)a_{m_1}(t)a_{m_2}(t) \nonumber \\
	&\times e^{i(E_n+E_{m_3}-E_{m_1}-E_{m_2})t},
\end{align}
where
\begin{align}
	V_{m_1,m_2}^{n,m_3}=\int {\rm d}x\, \phi_n^*(x)\phi_{m_3}^*(x)\phi_{m_1}(x)\phi_{m_2}(x).
\end{align}
Eq.~\eqref{eq:td_Anderson_modes} describes the excitation of new modes not already excited initially. Taking $a_m(t=t_I)$ as initial conditions, it is the dynamics built into Eq.~\eqref{eq:td_Anderson_modes} that governs scenario C while for scenario B $i\partial_t a_n=0$ holds for all $t> t_I$. The redistribution of mode excitations is determined by a subtle interplay between spatial overlap of the modes and their energies. In a disordered system modes with near degenerate eigenenergies must have small (spatial) overlap in order to suppress level repulsion. Two mechanisms should play a major role: the near resonant excitation of modes with center of mass outside the initial spread of the wavepacket at $t_I$ \cite{FlaKriSko2009,LapBodKri2010,FisKriSof2012}, and a non-resonant excitation of modes with center of mass close to the origin but with larger energies and localization lengths. The spreading of the momentum distribution [Fig.~\ref{fig:wfn_C} (b)] indicates the importance of the latter mechanism. 
\section{Sub-diffusive spreading and localization}\label{sec:macobs}
%
\begin{figure}[t]
	\includegraphics[width = 0.9\columnwidth]{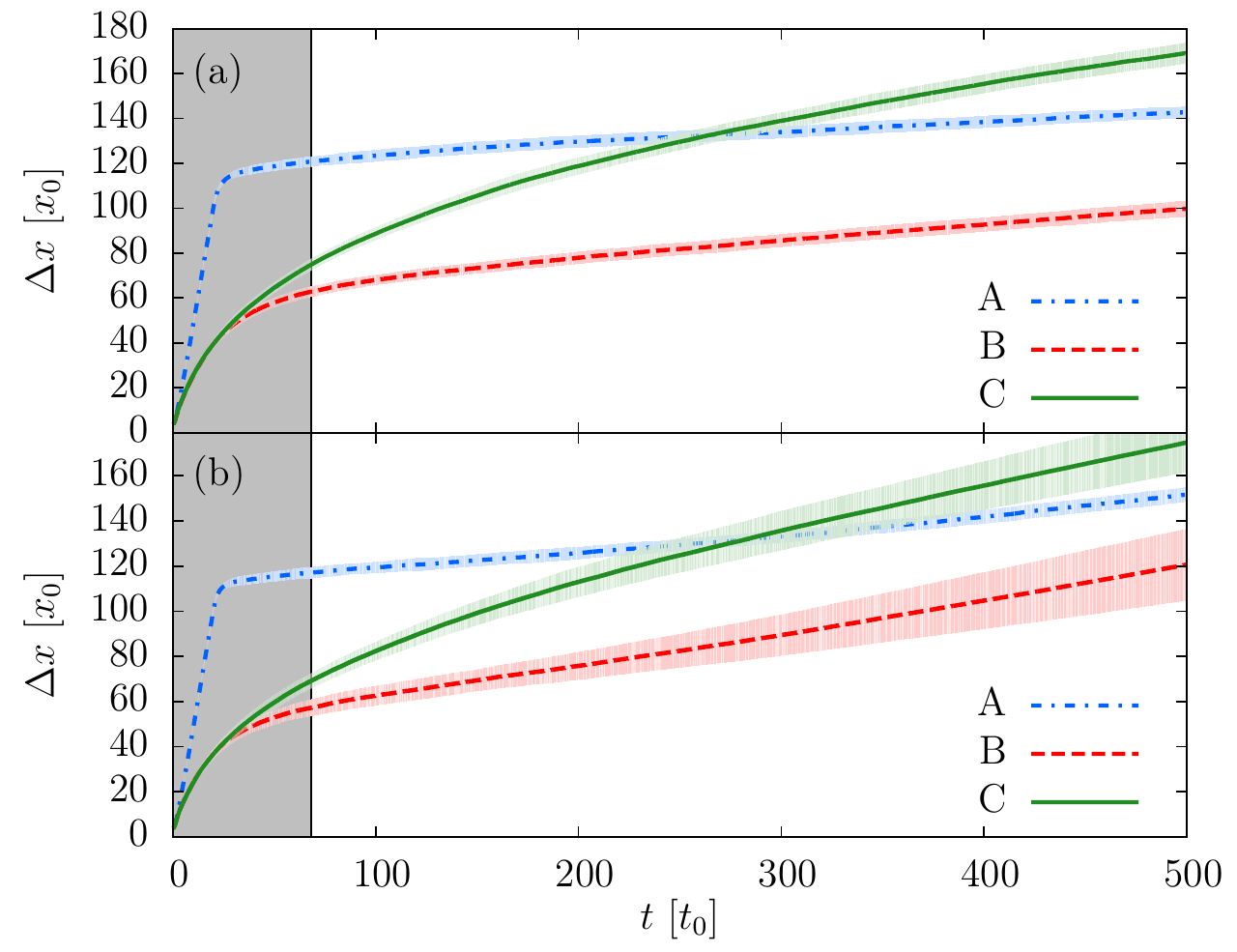}
	\caption{(Color online) The width $\Delta x (t)$ (Eq.~\ref{eq:delta_x}) as a function of time for the three scenarios for (a) the Gaussian correlated disorder, and (b) the speckle disorder. The colored shaded area around each curve represents one standard deviation resulting from the ensemble of $50$ disorder realizations each. The gray bar marks the time accessed experimentally \cite{BilJosZuo2008}.}
	\label{fig:delta_x}
\end{figure}
\begin{figure}[t]
	\includegraphics[width = 0.9\columnwidth]{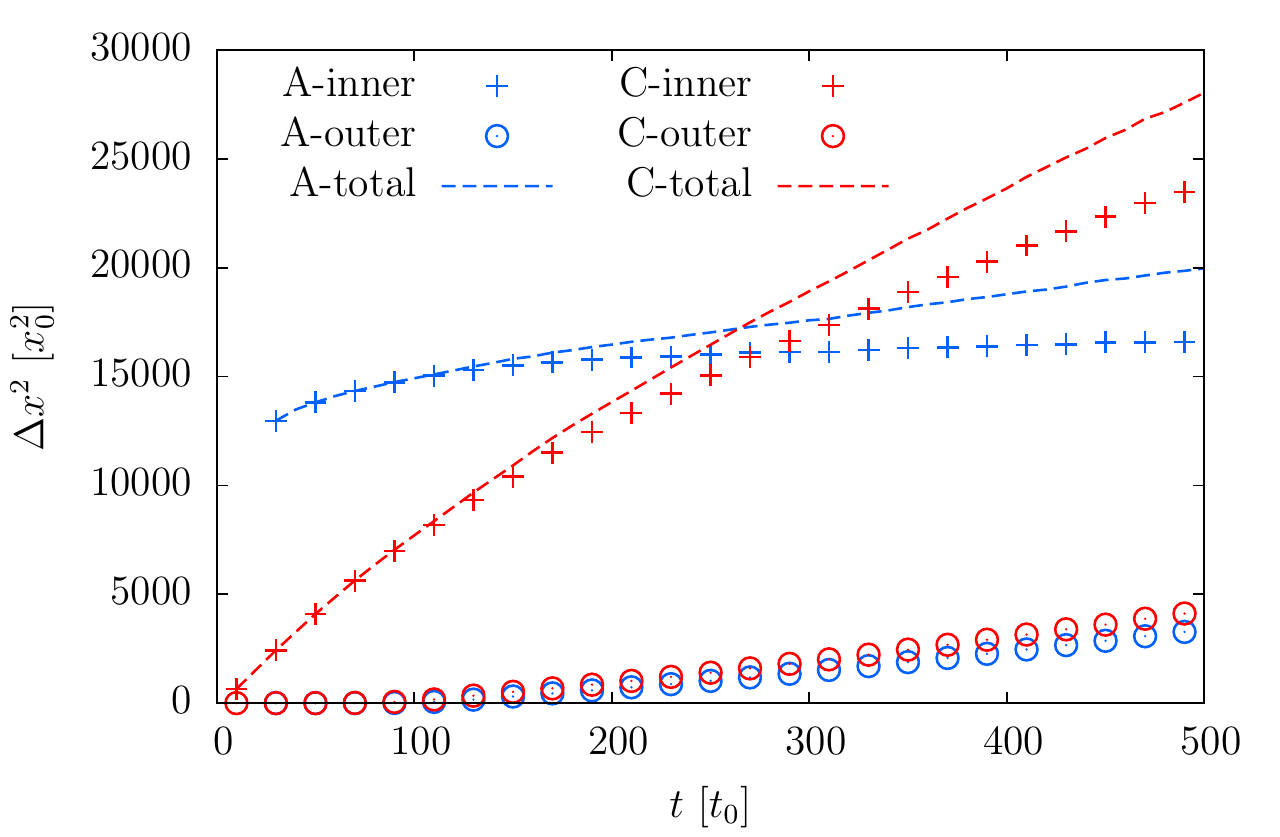}
	\caption{(Color online) Time dependent second moment $\Delta x^2 (t)$ (Eq.~\ref{eq:delta_x}) for scenario A and C for the Gaussian correlated disorder. The labels ``inner" and``outer" reflect the contributions to $\Delta x^2 (t)$ when only the contributions from the inner region $x \in \left[-3000:3000 \right] x_0$ or only from the outer region $\left|x\right|>3000 x_0$ are included. These partial second moments are additive $\left< \Delta x ^2 \right>_{\mathrm{inner}}+\left< \Delta x ^2 \right>_{\mathrm{outer}}=\left< \Delta x ^2 \right>_{\mathrm{total}}$.}
	\label{fig:delta_x_in_out}
\end{figure}
\begin{figure}[t]
	\includegraphics[width = 0.9\columnwidth]{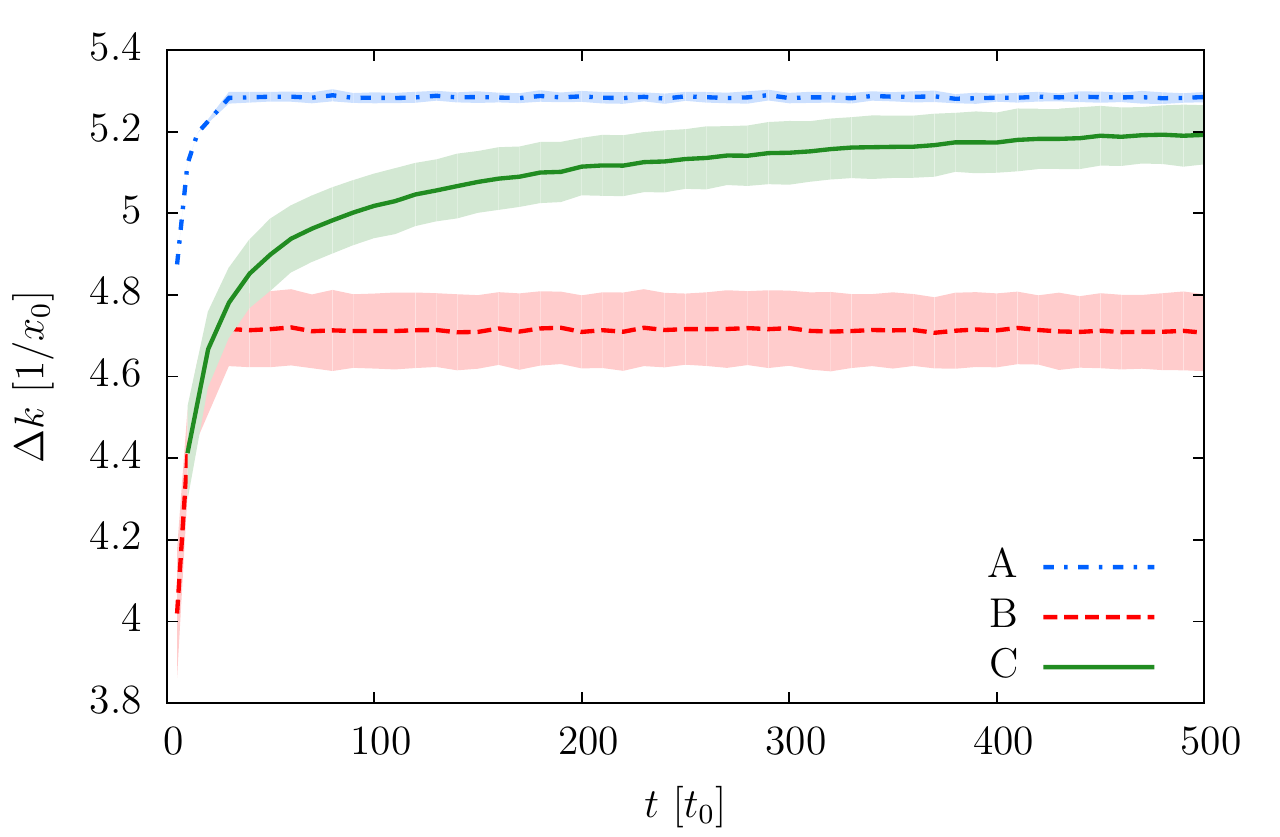}
	\caption{(Color online) The width $\Delta k(t)$ (Eq.~\ref{eq:delta_k}) as a function of time for the three scenarios for the Gaussian correlated disorder. The colored shaded area around each curve represents one standard deviation for an ensemble of $50$ disorder realizations.}
	\label{fig:delta_k}
\end{figure}
In this section we focus on observables that have been used in the past to characterize spreading or localization of the expanding BEC. The behavior of the width $\Delta x$ [Eq.~\eqref{eq:delta_x}] has gained considerable attention \cite{FlaKriSko2009,LapBodKri2010, FisKriSof2012, LucDeiTan2011}. For discrete systems described by a non-linear Schr\"odinger equation, sub-diffusive spreading, i.e. $\Delta x (t) \propto t^{\alpha}$ with $\alpha < 1/2$ has been predicted and observed. The exponent is predicted to be $\alpha=1/4$ if the interaction energy is larger than the mean level spacing within the wavepacket, and $\alpha=1/6$ if the interaction energy is smaller than the mean level spacing \cite{FlaKriSko2009,LapBodKri2010}. \\
These results for the discrete system cannot be directly transcribed to the continuum system investigated here. One important difference between the discrete system and the continuous system is that the energy spectrum of the discrete system is bounded both from below and from above while for the continuum system it is only bounded from below. Therefore, self-trapping, where the interaction energy is larger than the upper bound of the spectrum, cannot occur in the continuum system.\\
As $\Delta x (t)$ measures the overall width of the wavepacket, it is particularly sensitive to the ballistically moving tails. Due to the presence of these tails in the asymptotic densities, even in scenario A without the interaction present $(g=0)$ the width as a function of time, $\Delta x(t)$, grows, and does not saturate (Fig.~\ref{fig:delta_x}). To disentangle the contributions to the increase of $\Delta x ^2$ due to the spreading of the central part of the wavepacket from that for the quasi-ballistic growth of the tails we calculate the contributions to $\Delta x^2 (t)$ from both within the interval $x \in \left[-3000:3000 \right]x_0$ and outside of it (Fig.~\ref{fig:delta_x_in_out}). While in scenario A the central part of the wavepacket does not grow significantly after $t \approx 250 t_0$, indicating the onset of localization, the wavepacket spreads continuously for scenario C showing no sign of localization. The outer part containing the tails originating from the high momentum components, however, shows a similar increase for both scenarios and strongly contributes to the total spread of the wavepacket for later times. Scenario B shows qualitatively the same behavior as scenario A.\\
\begin{figure}[t]
	\includegraphics[width = 0.9\columnwidth]{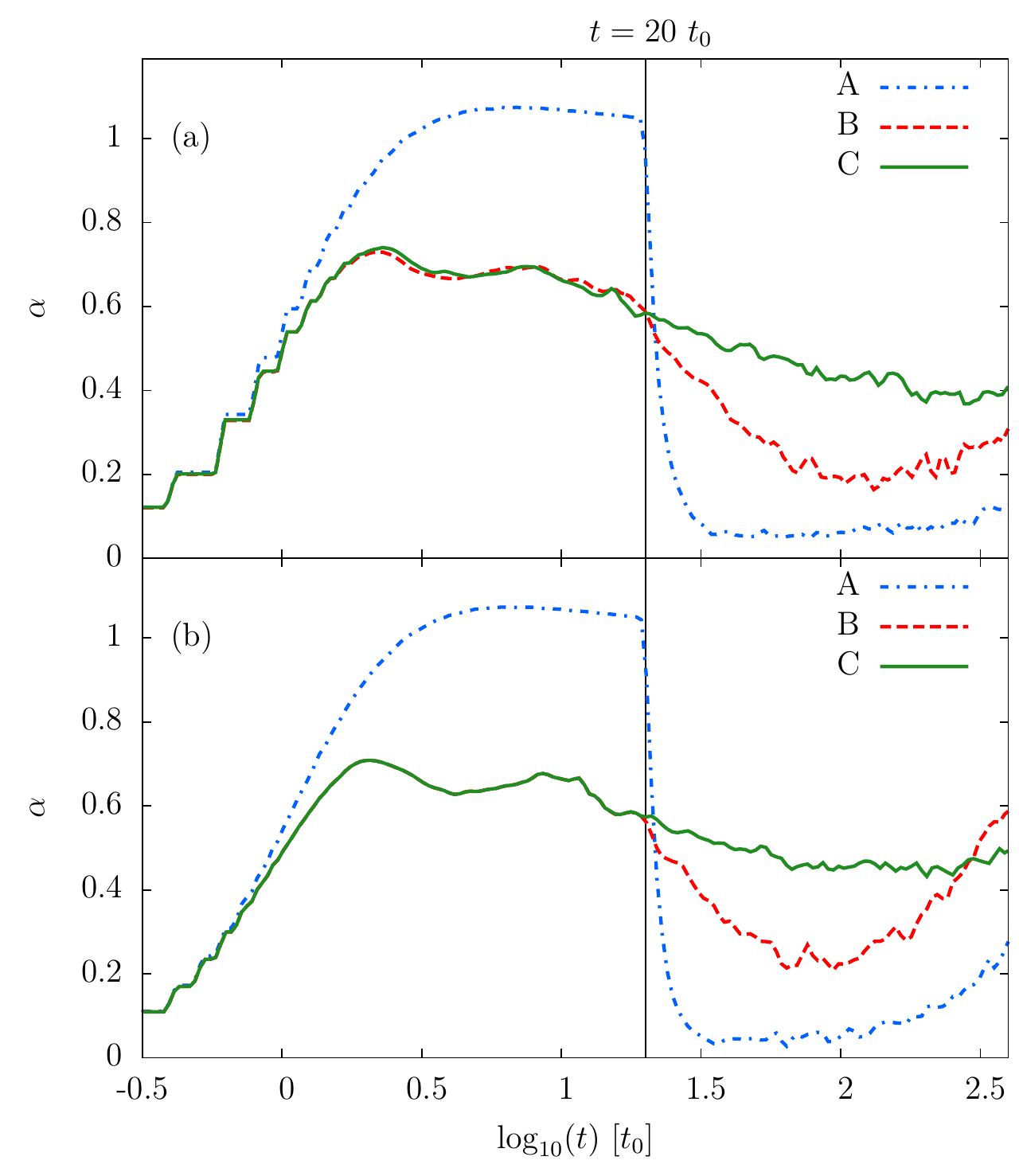}
	\caption{Time dependence of the exponent $\alpha(t)$ (Eq.~\ref{eq:alpha}) obtained from the results in Fig.~\ref{fig:delta_x} for (a) the Gaussian correlated disorder and (b) the speckle potential.}
	\label{fig:alpha}
\end{figure}
Interestingly, we observe in scenarios B and C for the speckle potential larger deviations of $\Delta x(t)$ for individual realizations of the disorder as indicated by the larger standard deviation [Fig.~\ref{fig:delta_x} (b)]. This is due to the fact that the peak values of the disorder potential and, consequently, the total energy fluctuates more strongly between individual realizations of the disorder for the speckle than for the Gaussian correlated disorder. Since in scenario A the wavepacket has already substantially spread when disorder is switched on, the standard deviation of $\Delta x$ is reduced in this case compared to that for scenarios B and C. For the width in momentum space $\Delta k(t)$ we observe, in contrast to the behavior of $\Delta x(t)$, for $g=0$ (scenarios A and B) a saturated momentum distribution which translates into $\Delta k(t)$ becoming stationary as a function of time (Fig.~\ref{fig:delta_k}). \\
In order to characterize the (sub)diffusive spreading and to compare with discrete models we extract estimates for the time-dependent exponents 
\begin{align}\label{eq:alpha}
	\alpha(t) = \frac{{\rm d}\log_{10}{\Delta x(t)}}{{\rm d}\log_{10}{t}}
\end{align}
for all scenarios (Fig.~\ref{fig:alpha}). Within the time interval accessible by our simulation which exceeds the observation time in the experiment \cite{BilJosZuo2008} by a factor 7, convergence to a well-defined exponent cannot (yet) be observed. For both scenarios A and B in which the long-time expansion proceeds in the absence of interactions, $\alpha(t)$ is still increasing for large $t$. Only for scenario C, $\alpha(t)$ is only weakly varying which may indicate an approach to a saturated value. If so, its value is, however, much higher than previously observed for discrete models and with $\alpha \approx 0.4 - 0.5$ close to the transition from the sub-diffusive to the diffusive regime.\\
The large values of $\alpha$ observed within all scenarios reflect the presence of fast quasi-ballistically moving tails absent in discrete models.  It should be emphasized, however, that for an unambiguous determination of the long-time limit $\alpha (t \rightarrow \infty)$ still much longer propagation times would be needed. These would, in turn, also require exceedingly large box sizes as the rapidly moving tails will eventually reach the wall of the box. A conclusive numerical test appears currently not yet computationally feasible. Generally, the extracted value of $\alpha$ cannot differentiate between the expansion of the outer region caused by the high-momentum tails and the expansion of the central region caused by particle-particle interactions (Fig.~\ref{fig:delta_x_in_out}). \\
As the ballistic expansion is generic for expanding BECs and not specific to the presence of particle-particle interactions, $\Delta x (t)$ appears not to be a suitable measure to probe the influence of interactions on Anderson localization of an expanding BEC. Since the long-time spread $\Delta x (t)$ is currently neither experimentally nor computationally accessible and, moreover, strongly influenced by the ballistic tails of the wavepacket, it is useful to focus on alternative and more local measures for localization that address the shape evolution on smaller length scales close to the center of $n(x,t)$.
\begin{figure}[t]
	\includegraphics[width = 0.9\columnwidth]{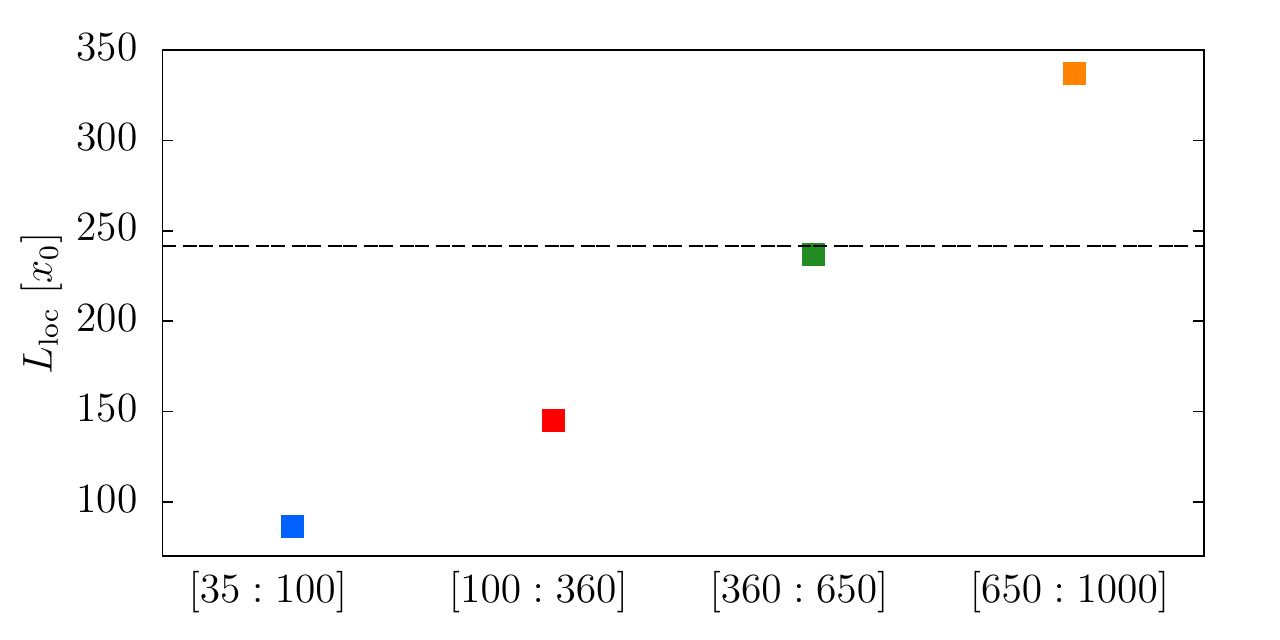}
	\caption{(Color online) Localization length $L_{{\rm loc}}$ for scenario B with a speckle potential evaluated by fitting the densities from Sec.~\ref{sec:sce} at $t=500 t_0$ to Eq.~\ref{eq:exp_loc_length} for different intervals in units of $x_0$. The horizontal dashed line shows the analytic prediction (Eq.~\ref{eq:loc_length_emax}).} 
	\label{fig:loc_length_diff_int}
\end{figure}
\begin{figure}[t]
	\includegraphics[width = 0.9\columnwidth]{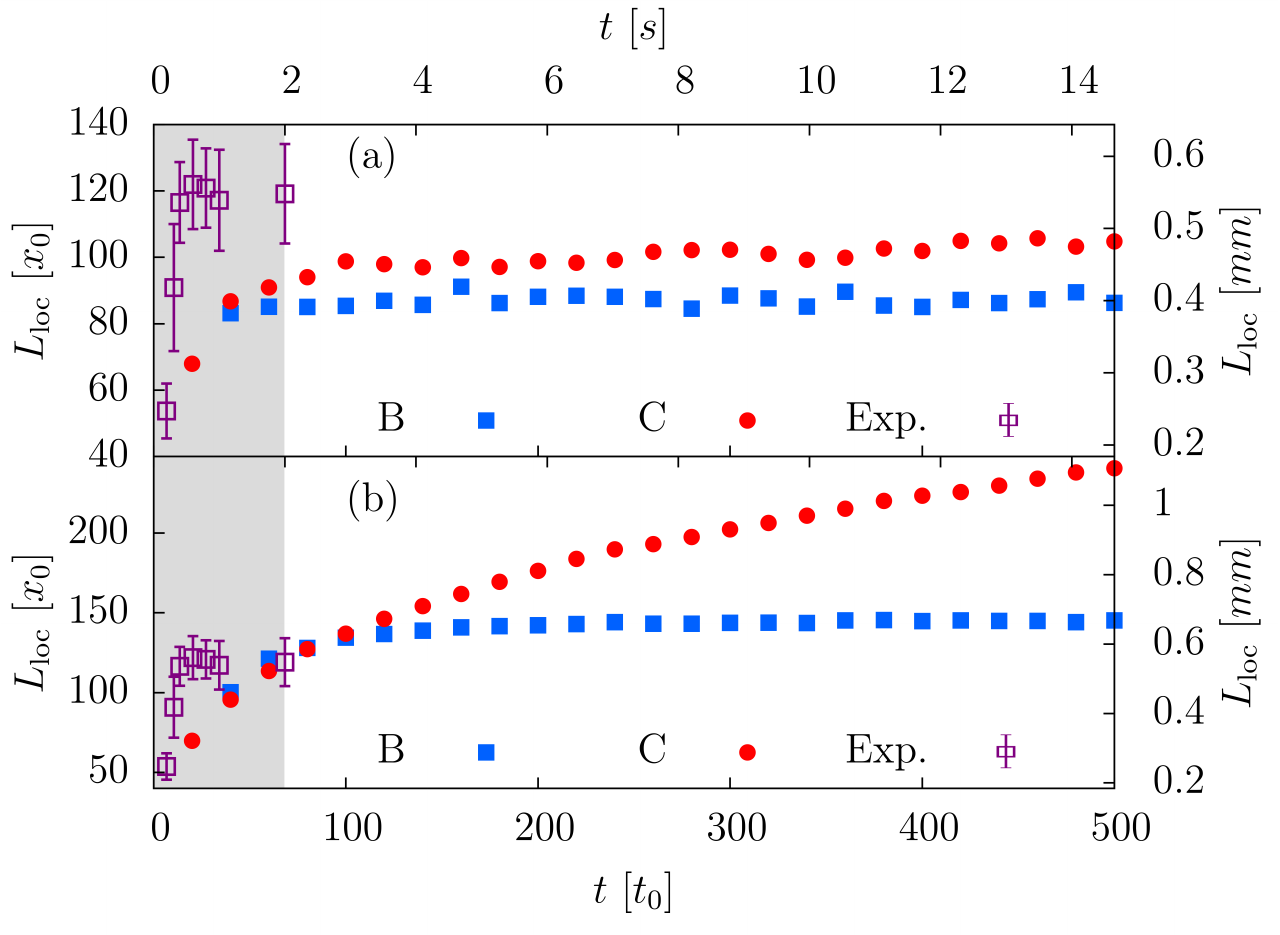}
	\caption{(Color online) Localization length $L_{{\rm loc}}$ for the speckle potential evaluated by fitting the densities from Sec.~\ref{sec:sce} to Eq.~\ref{eq:exp_loc_length} in the interval (a) $x\in[35:100]x_0$ and (b)  $x\in[100:360]x_0$. The gray shaded area indicates the time accessed in the experiment. The experimental values for $L_{\rm loc}$ are taken from Ref.~\cite{BilJosZuo2008}. } 
	\label{fig:loc_length}
\end{figure}
While $n(x,t)$ globally does not display a simple exponential shape 
\begin{align}\label{eq:exp_loc_length}
	n(x,t)\sim \exp \left( -2 \left| x\right| / L_{\rm loc} \right),
\end{align}
near the center of the distribution one can extract an effective exponential slope within a spatially restricted interval. To this end, we fit the densities in Sec.~\ref{sec:sce} to Eq.~\eqref{eq:exp_loc_length} in different intervals. We address in the following only the scenarios B and C because the shape of the central peak of the density of these two is similar and considerably different from scenario A. Subsequently we focus on speckle disorder for comparison with the experimental results of Ref.~\cite{BilJosZuo2008}. The results for Gaussian disorder are, however, very similar. In previous works \cite{BilJosZuo2008,SanCleLug2008} it was estimated that the effective localization length $L_{\rm loc}$ for scenario A is given by 
\begin{align}\label{eq:loc_length_emax}
	L_{\rm loc}\simeq \lambda \left(E_{\mathrm{max}}\right).
\end{align}
with $\lambda$ given by Eq.~\eqref{eq:gamma} and the maximum energy $E=E_{\mathrm{max}}=1/2\xi^2$  after the free expansion until $t=t_D=t_I$.
Since within scenario B the momentum and energy distribution reached at $t=t_I$ differs from that of scenario A it is a priori not obvious that Eq.~\eqref{eq:loc_length_emax} should hold for scenario B as well. Nevertheless, since the high-momentum cut-off after the early expansion remains essentially unchanged by the friction of the wavepacket in the disorder potential, Eq.~\eqref{eq:loc_length_emax} provides still a useful estimate for $L_{\rm loc}$ for scenario B. 
Since the shape of the numerically simulated central peak of the density is not characterized by a single exponential, the fitted value for $L_{\rm loc}$ sensitively depends on the interval of x coordinates included in the fit. We probe this dependence for scenario B for which the density distribution near the maximum converges (see Fig.~\ref{fig:wfn_B}). The inverse slope $L_{\rm loc}$ increases by more than a factor 3 when its interval is shifted from close to the central peak $[35:100]$ in units of $x_0$ further out to the wings $[650:1000]x_0$ (Fig.~\ref{fig:loc_length_diff_int}). The approximately predicted value of $L_{\rm loc} \approx 242 x_0$ [Eq.~\eqref{eq:loc_length_emax}] is found near an intermediate interval  $[360:650]x_0$.\\
In view of this variation of the local slope a comparison with the experimental data of Ref.~\cite{BilJosZuo2008} is not unambiguous. In Fig.~\ref{fig:loc_length} we compare the experimental data for the evolution of $L_{\rm loc}$ as a function of $t$ with our numerical prediction for scenarios B and C in two different extraction intervals, close to the peak $[35:100] x_0$ [Fig.~\ref{fig:loc_length}~(a)] and at larger distance $[100:360] x_0$ [Fig.~\ref{fig:loc_length}~(b)]. For the latter case, we find reasonable agreement with experimental data up to the largest times observed in the experiment ($t=68 t_0$ or 2 seconds). The significance of this agreement should, however, be viewed with caution. Apart from the uncertainty in the interval over which the experimental decay of the density has been determined the experimental data does not cover the following evolution where scenarios B and C begin to diverge ($t > 100 t_0$) and where scenario B yields, indeed, a stationary value of $L_{\rm loc}$ while scenario C does not [Fig.~\ref{fig:loc_length} (a) and (b)]. The particle-particle interactions included in scenario C render the slope of the central peak time-dependent yielding a monotonically growing $L_{\rm loc}$.
\begin{figure}[t]
	\includegraphics[width = 0.9\columnwidth]{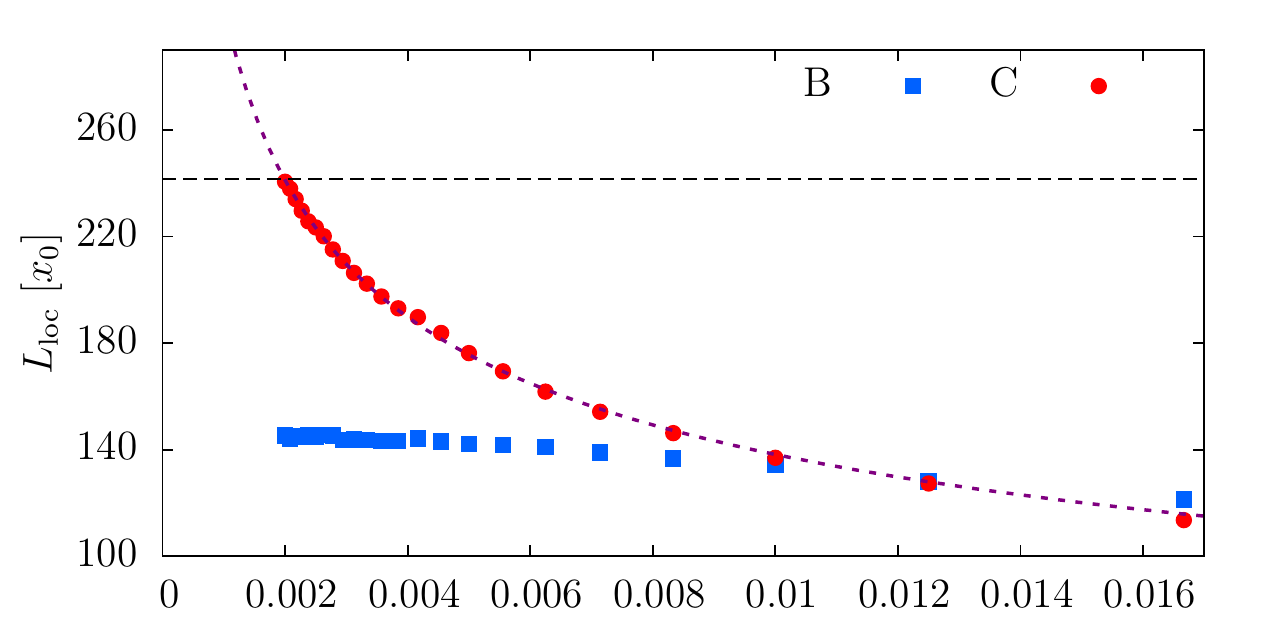}
	\caption{(Color online) Same as Fig.~\ref{fig:loc_length}, but plotted as function of inverse time $1/t$ for the interval $x\in[100:360]x_0$. The horizontal dashed line shows the analytic prediction (Eq.~\ref{eq:loc_length_emax}). The dashed purple line following the points of scenario C is a fit to a function $f(1/t)=a(1/t)^b$ with $b \approx 0.35$.} 
	\label{fig:loc_length_inv_t}
\end{figure}
Extrapolation on a $t^{-1}$ plot (Fig.~\ref{fig:loc_length_inv_t}) suggests that $L_{\rm loc}(t)$ may diverge. To test this assumption we fit the data points for scenario C to a function $f(1/t)=a\left(1/t\right)^b$ for the extraction interval $x \in \left[100:360\right] x_0$ (purple dashed lines in Fig.~\ref{fig:loc_length_inv_t}). We find a finite value of $b \approx 0.35$ suggesting that $L_{\rm loc}$ does, indeed, diverge. As for $\Delta x (t)$, also for $L_{\rm loc}(t)$ a definite conclusion on the asymptotic behavior cannot be drawn from the present simulation. The point to be noted is that the apparent divergences for $\Delta x (t)$ and $L_{\rm loc}(t)$ occur on different length scales and are due to fundamentally different processes. While the continued growth of $\Delta x (t)$ is mostly unrelated to particle-particle interactions in the late phase of the expansion but originates from excitation of energetically high-lying Anderson modes with very large (or even diverging) localization length in the early phase of the expansion, the continued growth of $L_{\rm loc}$ (inverse slope of the central peak) is a true interaction effect at late times. This observation strongly suggests that particle-particle interactions, indeed, destroy Anderson localization. We emphasize that the difference between the non-interacting and interacting case (scenario B and scenario C) becomes visible only after relatively long times ($t=150 t_0$ which corresponds to 4.5s) beyond the observation time in \cite{BilJosZuo2008}. Observation of this effect in future experiments may pose a challenge. The present results provide quantitative predictions for the time scale on which the role of interactions on Anderson localization in the expansion of a BEC become experimentally observable.
\section{Summary and conclusions}\label{sec:sum}
We have investigated the expansion of a one-dimensional (1D) Bose-Einstein condensate (BEC) in Gaussian correlated  as well as speckle disorder closely following the experimental setup in Ref.~\cite{BilJosZuo2008}. For the theoretical and numerical description of the system we have used the Gross-Pitaevskii equation (GPE). We addressed the role of interactions (i.e.~of nonlinearities) on Anderson localization by studying three different scenarios that allow to disentangle effects due to disorder and interactions during the expansion process. The conversion of interaction energy into kinetic energy in the early phase of the expansion results in a broad energy distribution of the wavepacket and fast moving tails. In the presence of a disorder landscape this broad excitation spectrum corresponds to the excitation of energetically high-lying Anderson modes with large (or even diverging) localization lengths. The presence of these ballistic components renders one of the frequently involved indicators for Anderson localization, the saturation of the width $\Delta x$ of a wavepacket as a function of time, largely inapplicable, even in the absence of interactions at large times. When both disorder and interactions are simultaneously present, $\Delta x$ grows even faster with time. In this case also the momentum distribution continuously grows indicating that new Anderson modes with larger energies and localization lengths are excited due to particle-particle interactions. \\
We have investigated whether the width obeys a power law $\Delta x\propto t^\alpha$ with $\alpha<1/2$ predicted analytically and verified numerically for the discrete system \cite{FlaKriSko2009,LapBodKri2010}. Our numerical results show that even in the absence of interactions, $\alpha (t)$ still grows at late times rendering convergence to a definite exponent $\alpha < 1/2$ for subdiffusive spreading unlikely due to the rapidly expanding tails originating from high-lying Anderson modes. These have not been observed in previous works and may be absent in discrete systems. When both disorder and interactions are present, the time dependence of $\alpha(t)$ becomes markedly weaker. The saturation value, if it exists, is however much higher than for discrete systems and close to the diffusive limit $\alpha \approx 0.5$. On a shorter distance scale the inverse slope of the approximately exponential decay of the central peak in the real-space particle density can provide a useful local measure for the localization length $L_{\rm loc}$. In the absence of interactions at late times the numerically extracted values for $L_{\rm loc}$ become (approximately) time-independent and are of the order of the analytic prediction for the localization length for the Anderson modes with energies close to the maximum energy realized in the early phase of the expansion. However, with interactions present, the inverse slope and, thus, $L_{\rm loc}$ continues to increase up to long times. We show that the extracted values of $L_{\rm loc}$ is strongly dependent on the interval over which the slope is measured. Differences between the interacting and the non-interacting case become significant only after $t=150t_0$ which corresponds to $4.5$s with the parameters from the experiment \cite{BilJosZuo2008}. This time is substantially longer than the previously accessed time and sets a benchmark for future experiments that can address the role of interactions on Anderson localization in ultra-cold atoms. The present numerical results suggest that, indeed, interactions destroy localization also on this length and time scale.
\subsection*{Acknowledgements}
We acknowledge fruitful discussions with Winfried Auzinger, Dmitri Krimer, and Peter Schlagheck. Calculations have been performed on the Vienna Scientific Cluster (VSC3). This work has been supported by the WWTF project MA14-002, the FWF through FWF-SFB042-VICOM, FWF-SFB049-NEXTlite, FWF Doctoral College Solids4Fun (W1243) as well as the IMPRS-APS.
\appendix
%
\section{Numerical method for time propagation }\label{App:numerics}
For the numerical time integration of the GPE which belongs to the class of nonlinear Schr{\"o}dinger equations of the type
\begin{equation}\label{de1}
\mathrm{i} \partial_t \psi(t) =  A\psi(t) + B[\psi(t)],\quad t>t_0,
\end{equation}
we employ high-order adaptive multiplicative operator splitting methods.
These $s$-stage exponential splitting methods for the integration of Eq.~(\ref{de1})
use multiplicative combinations of the partial flows $ \phi_A(t,\psi) $ and $ \phi_B(t,\psi) $.
For a single step $ (0,\psi_0) \mapsto (dt,\psi_1) $ with timestep $ t=\text{d}t$, this reads
\begin{align}\label{splitting1}
	\psi_{1} &:= \mathcal{S}(\text{d}t,\psi_0) \nonumber \\
	&= \phi_B(b_s \text{d}t,\cdot) \circ \phi_A(a_s \text{d}t,\cdot)\nonumber \\
	&\circ \ldots \circ
	\phi_B(b_1 \text{d}t,\cdot) \circ \phi_A(a_1 \text{d}t,\psi_0),
\end{align}
where the coefficients $ a_j,b_j,\, j=1 \ldots s $ are determined according to the requirement
that a preset order of consistency is obtained \cite{HaiLubWan2002}. The dot in the brackets symbolizes that during each application of a new partial flow propagator $\phi$ the most current realization of the state vector is used. $\phi_B\circ \phi_A$ represents the subsequent application of the partial flow $\phi_B$ after the partial flow $\phi_A$.
The system in our case contains two vector fields, the kinetic energy part and the potential energy consisting of both the nonlinearity and the disorder potential, of different stiffness. The stiff flow is associated with the kinetic energy. If these are treated separately the resulting subproblems can typically be integrated with more efficient schemes.
For the GPE the kinetic part can be integrated efficiently after (pseudo)spectral space discretization by exponentiation of
a diagonal matrix while the nonlinear part allows an exact integration of the resulting
ordinary differential equation in real space. Thus, the numerical effort effectively reduces
to transformations between real and Fourier space which can be implemented with
low cost also in a parallel environment. In the present case we use an equidistant grid and employ fast Fourier transforms within the package FFTW \cite{FriJoh2005}.\\
The efficiency of the time discretization can be improved if high-order time propagators are employed.
These provide a more efficient approximation if the exact solution is sufficiently regular. In the present case we have used a fourth-order propagator as introduced in Ref.~\cite{AuzHofKet2017,splithp}.\\
Adaptive choice of the time steps yields a further reduction in the computational effort.
\begin{figure}[t]
	\includegraphics[width = 0.9\columnwidth]{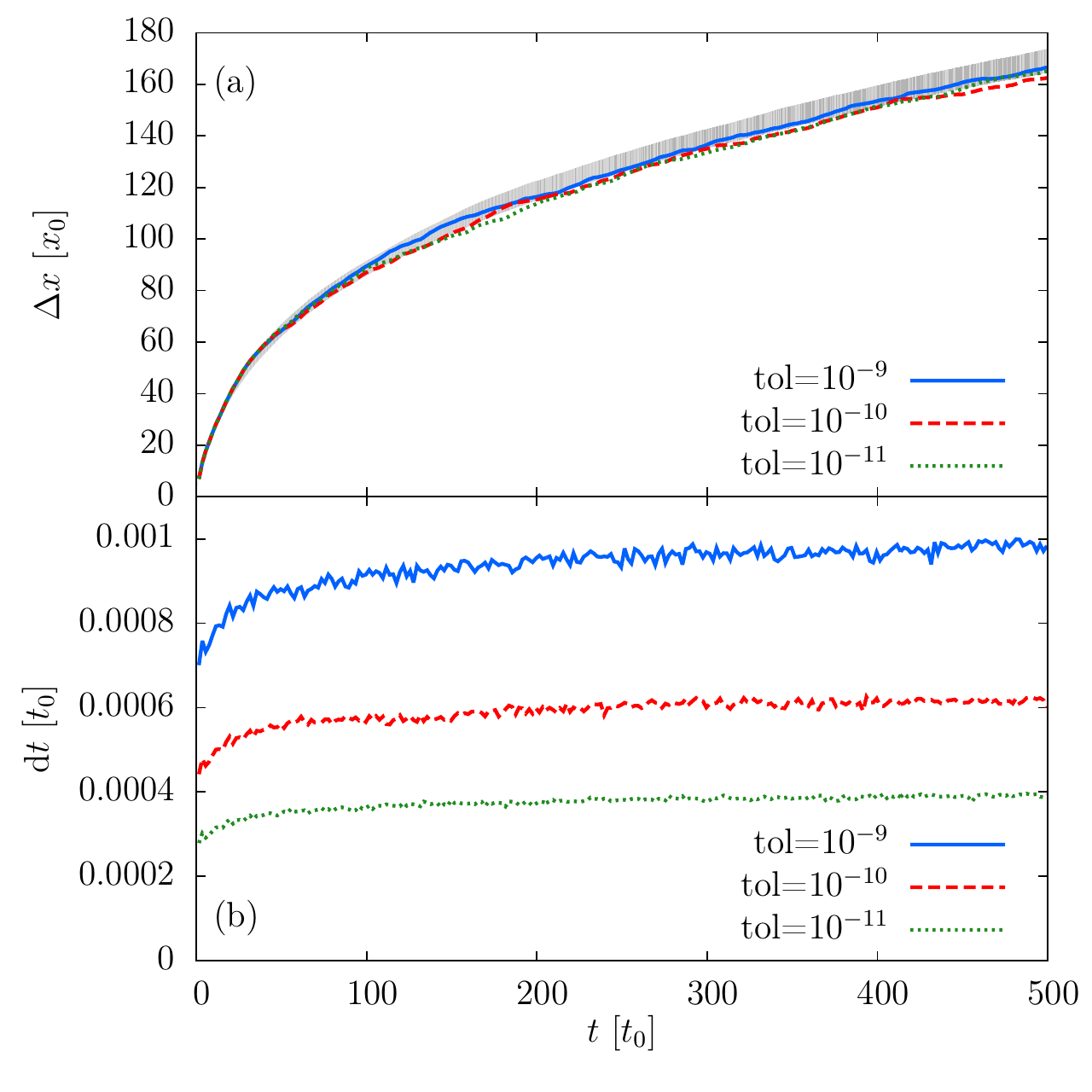}
	\caption{(Color online) (a) The width $\Delta x(t)$ (Eq.~\ref{eq:delta_x}) as a function of time for different tolerances ${\rm tol}$ that result in different average time steps (Eq.~\ref{eq:step-selct}) for a single realization of a Gaussian correlated disorder. The gray colored area represents one standard deviation around the mean value resulting from an ensemble of 50 disorder realizations. (b) The corresponding time step for different tolerances.}
	\label{fig:conv_tol}
\end{figure}
As a basis for time-step adaption asymptotically correct a posteriori estimators of
the local time-stepping error are required. We can choose from four classes of local error estimators
which have different advantages depending on the context in which they are applied (for a review see e.g.~Ref.~\cite{AuzHofKet2017}).
Embedded pairs of splitting formulae have been introduced in
\cite{KocNeuTha2013} and are based on reusing a number of evaluations from
the basic integrator. In our simulations, we have used an embedded pair of orders 4(5) referred to as \texttt{Emb 5/4 AK (ii)} in the collection \cite{splithp}.\\
Based on a local error estimator the step size is adapted such that the prescribed tolerance
is expected to be satisfied in the following step. If $\text{d}t_{\mathrm{old}}$ denotes the
current step-size the next step-size $\text{d}t_{\mathrm{new}}$ in an order $p$ method is predicted as (see \cite{PreFlaTeu1988})
\begin{equation}
\label{eq:step-selct}
\text{d}t_{\mathrm{new}} = dt_{\mathrm{old}} \cdot \min\Big\{\alpha_{\mathrm{max}},\max\Big\{\alpha_{\mathrm{min}},
\Big(\alpha\,\frac{\mathrm{tol}}{\mathcal{P}(\text{d}t_{\mathrm{old}})}\Big)^{\frac{1}{p+1}}\,\Big\}\Big\},
\end{equation}
where $ \alpha = 0.8 $, $ \alpha_{\mathrm{min}} = 0.25 $, $ \alpha_{\mathrm{max}} = 4.0 $,
and $\mathcal{P}$ denotes an estimate of the local error.
This strategy incorporates safety factors to avoid an oscillating and unstable behavior.\\
In Fig.~\ref{fig:conv_tol} (b) we present the adaptive time step as a function of time for a disorder realization for different tolerances. For the calculations in the present work we use a tolerance of $\mathrm{tol}=10^{-10}$ corresponding to an average time step of $5.5\times 10^{-4}t_0$. For this time step the energy is conserved up to ten significant digits which is comparable to the results using a tolerance of $\mathrm{tol}=10^{-11}$. For speckle potentials the time step is about a factor of two larger for the same error tolerance than for Gaussian disorder potentials. The error in energy conservation is of the same order. Note that the present propagator allows for a time step which is $\approx 10^2$ larger than the one required by the leap frog propagator used previously \cite{BreColLud2011,BreLodStr2012}. Only through this speed up long time propagations as presented in this paper become possible. For the three different tolerance levels shown in Fig.~\ref{fig:conv_tol}~(a) the width of the wavepacket differs from each other by much less than one standard deviation of the ensemble average (see also the discussion in App.~\ref{App:convergence}).
\section{Convergence tests}\label{App:convergence}
\begin{figure}[t]
	\includegraphics[width = 0.9\columnwidth]{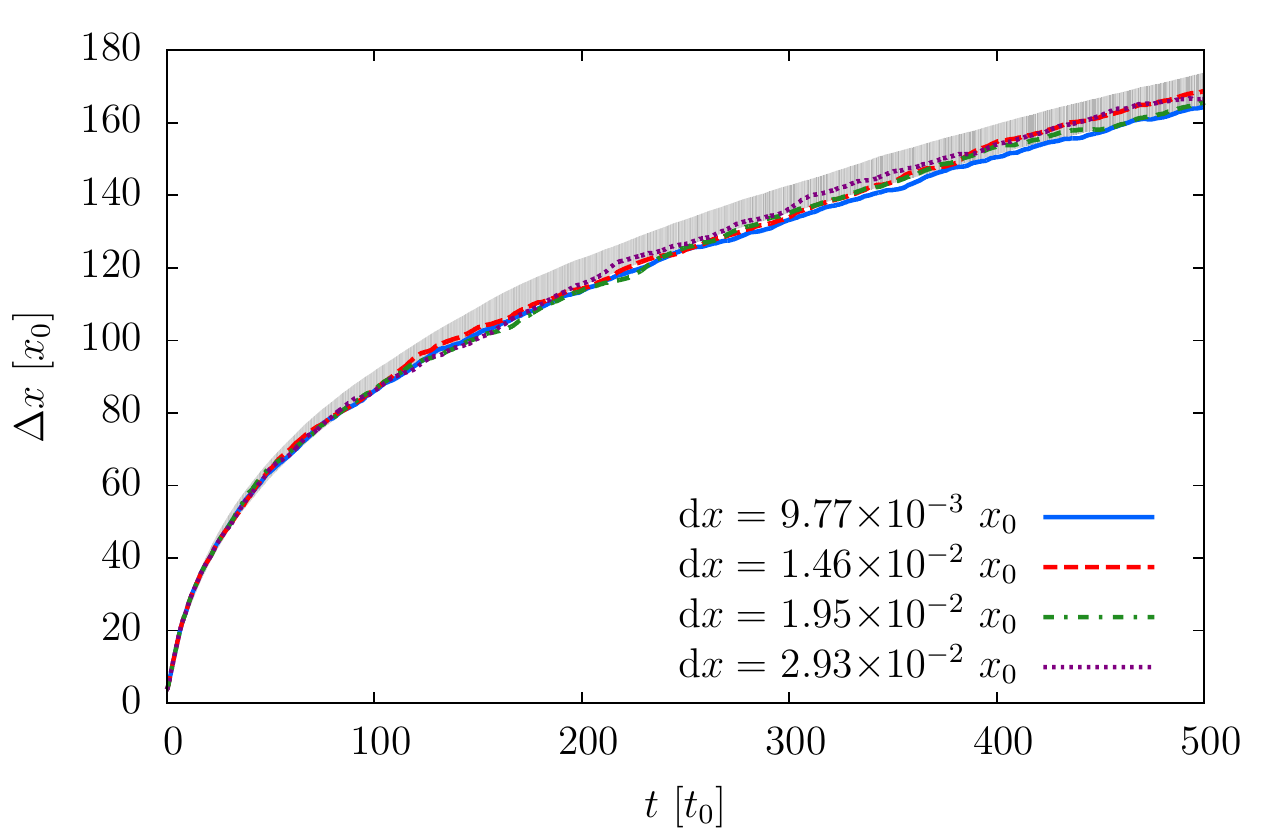}
	\caption{(Color online) The width $\Delta x(t)$ (Eq.~\ref{eq:delta_x}) as a function of time for different space discretizations. The numerical values are given up to the third significant digit. The gray colored area represents one standard deviation around the mean value resulting from an ensemble of 50 disorder realizations of the Gaussian correlated potential.}
	\label{fig:conv_space_dis}
\end{figure}

\begin{figure}[t]
	\includegraphics[width = 0.9\columnwidth]{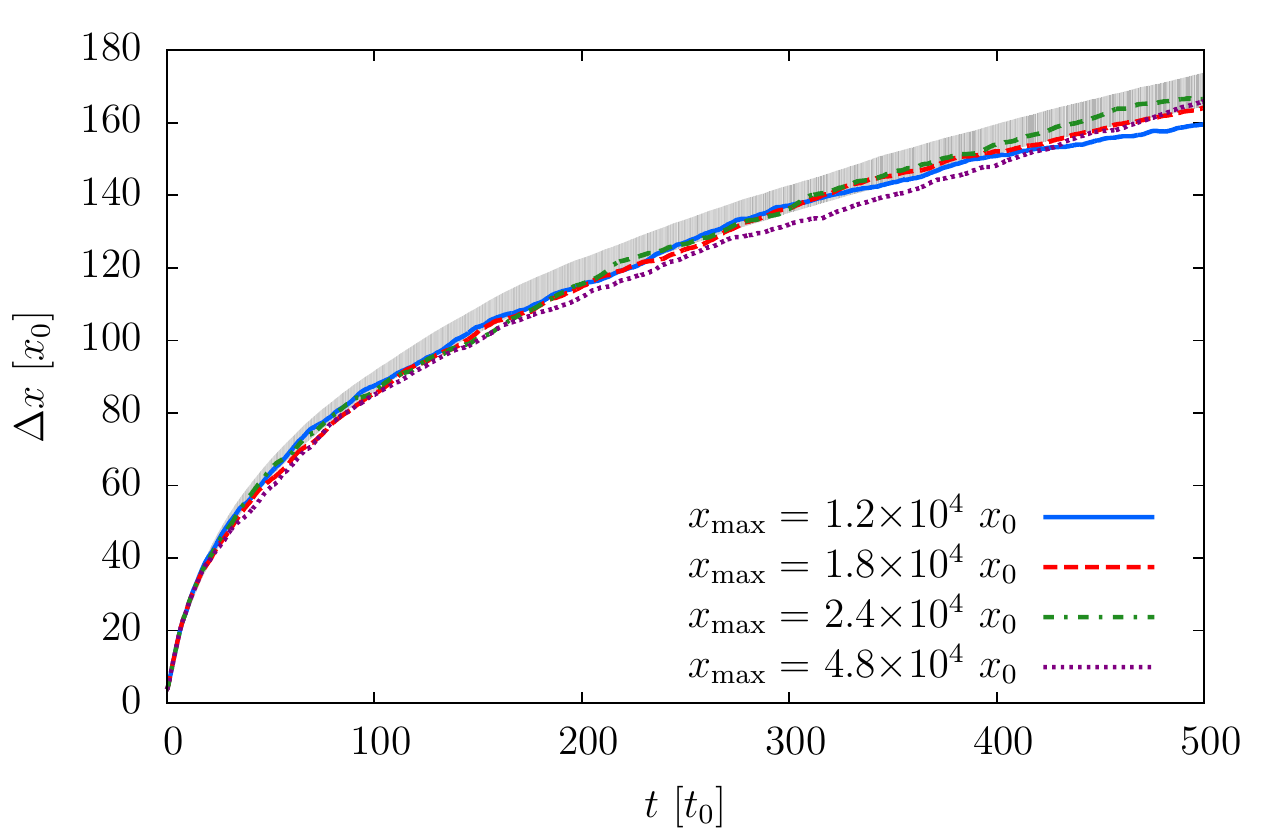}
	\caption{(Color online) (a) The width $\Delta x(t)$ (Eq.~\ref{eq:delta_x}) as a function of time for different box sizes. The box spans $[-x_{\mathrm{max}}:x_{\mathrm{max}}]$.  The light colored area represents one standard deviation around the mean value resulting from an ensemble of 50 disorder realizations }
	\label{fig:conv_box_size}
\end{figure}

In this section we discuss the convergence of our results with respect to spatial discretization and box size. The results presented here correspond to individual realizations of the disorder and do not contain any ensemble averaging. We show representative results only for scenario C and the Gaussian disorder potential as it is the numerically most challenging one. The results for speckle potentials are similar. Perfect agreement of each individual wavefunction for different error tolerances, spatial discretizations, and box sizes cannot be expected because of the underlying chaoticity of the GPE \cite{BreColLud2011}. However, we regard the results as converged if the differences are smaller than the variance within the ensemble of disorder realizations. \\
Comparing different spatial discretizations we observe that they agree very well within the error bars (one standard deviation) of the ensemble average (Fig.~\ref{fig:conv_space_dis}). For the results presented in the main text we use a spatial discretization of $\mathrm{d}x \approx 0.0293~x_0$. 
Comparing different box sizes $[-x_{\mathrm{max}}:x_{\mathrm{max}}]$ we see a rather large deviation after long times for the smallest box (Fig.~\ref{fig:conv_box_size}). All larger boxes give results within the standard deviation of the ensemble average. The influence of reflections off the walls at $-x_{\mathrm{max}}$, $x_{\mathrm{max}}$ are thus negligible. For the calculations in this work we use $x_{\mathrm{max}}=2.4\times 10^{4}$. With the spatial discretization given above this corresponds to a grid of 1638400 points.
\section{Analytical description of localized densities}\label{App:theory}
In this appendix we briefly review the key ingredients required to calculate the asymptotic densities (Eq.~\ref{eq:den_analyt}). The density of an Anderson localized state with fixed energy $E$ is given by the integral \cite{Ber1974, Gog1976}
\begin{align}
	n( \lambda(E), x)&= \frac{\pi^2\gamma(E)}{8}\int_0^{\infty} {\rm d}u\, u\,\sinh{(\pi u)}
	\left [\frac{1+u^2}{1+\cosh{(\pi u)}}\right]^2 \nonumber \\
	&\times \exp{[-(1+u^2)\gamma(E)|x|/2]},
\end{align}
with
\begin{align}
	\gamma(E)=\lambda^{-1}(E)=\frac{\sqrt{2 \pi}}{4 E}\tilde{C}\left(2 \sqrt{2 E}\right).
\end{align}
The spectral function 
\begin{align}
	A(k,E) = \frac{(-1/\pi) \Sigma''(k,E)}{\left[E-\frac{p^2}{2}-\Sigma'(k,E)\right]^2+\Sigma''(k,E)^2}
\end{align}
gives the probability amplitude to scatter into a state with energy $E$ at momentum $k$. The self-energy $\Sigma$ can be calculated in first order Born approximation yielding for the imaginary part
\begin{align}
	\Sigma''(k,E)=-\frac{\sqrt{2 \pi}}{2 \sqrt{2 E}}\left[\tilde C(\sqrt{2 E}-k) + \tilde C( \sqrt{2 E}+k)\right].
\end{align}
The real part $\Sigma '$ describing the deviation from the dispersion relation of a free particle is assumed to be small \cite{PirLugBou2011} and is neglected. 
%

\clearpage

%

\end{document}